\documentclass[preprintnumbers,amsmath,amssymb,superscriptaddress]{revtex4}
\usepackage{bm}
\usepackage{graphicx}
\usepackage{dcolumn}
\usepackage{xcolor}
\usepackage{hyperref}
\usepackage{times}
\usepackage{dcolumn}% Align table columns on decimal point
\usepackage{bm}% bold math
\usepackage{dsfont}
\usepackage{mathrsfs}
\usepackage{amsmath}
\usepackage{amsthm}
\usepackage{braket}

\newcommand{\Rmnum}[1]{\expandafter\@slowromancap\romannumeral #1@}

\begin{document}

\title{Efficient discrimination schemes for unextendible product bases with strong quantum nonlocality}

\author{Qiqi Feng}
\affiliation{School of Mathematics and Science, Hebei GEO University, Shijiazhuang 052161, China}

\author{Huaqi Zhou}
\email{zhouhuaqilc@163.com}
\affiliation{School of Mathematics and Science, Hebei GEO University, Shijiazhuang 052161, China}
\affiliation{Hebei Province Key Laboratory of Intelligent Sensing and Data Processing for Geo-environment, Hebei GEO University, Shijiazhuang 052161, China}

\author{Limin Gao}
\email{gaoliminabc@163.com}

\affiliation{School of Mathematics and Science, Hebei GEO University, Shijiazhuang 052161, China}
\affiliation{Hebei Province Key Laboratory of Intelligent Sensing and Data Processing for Geo-environment, Hebei GEO University, Shijiazhuang 052161, China}

\begin{abstract}
Entanglement is a central resource in quantum information science, and it is therefore important to design local discrimination protocols that minimize entanglement cost. In this paper, we propose several entanglement-assisted discrimination schemes for the local discrimination of a representative strongly nonlocal unextendible product basis (UPB) in a \(3\otimes 3\otimes 3\) system. By exploiting the structure of the UPB and the properties of maximally entangled resources, we generalize the protocols to a family of strongly nonlocal UPBs in \(d\otimes d\otimes d\) systems. In particular, we show that these UPBs can be perfectly distinguished using two bipartite maximally entangled states, distributed between different pairs of parties, without employing quantum teleportation. We further compare the total supplied and average consumed entanglement under a clearly specified accounting convention. The results demonstrate that avoiding teleportation can reduce the required entanglement in suitable resource-allocation scenarios and clarify the operational role of low-dimensional maximally entangled resources.
\end{abstract}

\maketitle

\section{Introduction}
Quantum state discrimination is one of the fundamental problems in quantum information theory. Beyond its role in information retrieval and state identification, it also underlies a broad range of tasks, including quantum secret sharing, quantum data hiding, and secure quantum communication \cite{Eggeling02,DiVincenzo2002,matthews2009,lo1999,rahaman2015,markham2008,Guo01,Hsu05,JWang17}. Accordingly, given a set of mutually orthogonal states, a key question is how to identify every state from this set.

For any set of mutually orthogonal pure quantum states in a multipartite system, perfect discrimination can always be achieved by global measurements \cite{bennett1999}. In contrast, under the restriction of local operations and classical communication (LOCC), spatially separated parties are restricted to local quantum operations coordinated by classical communication, where the operations may be adaptive \cite{bennett1999,bennett1999upb,divincenzo2003upb,rinaldis2004,ghosh2001}. Under this restriction, even orthogonal product states may become locally indistinguishable \cite{yang2015characterizing,halder2018several,xu2016quantum,wang2017local,zhang2017construction}.

In 1999, Bennett \emph{et al.}~\cite{bennett1999} first demonstrated ``quantum nonlocality without entanglement'' by constructing a complete orthogonal product basis in a
\(3\otimes3\) system. Although these states are perfectly distinguishable by global measurements, they cannot be perfectly distinguished by LOCC. This phenomenon reveals intrinsic limitations of LOCC-based discrimination, motivates systematic studies of local distinguishability, and has stimulated extensive research on entanglement-assisted discrimination. Since then, substantial progress has been made \cite{Walgate02,ZGY25,HGY25,HGY24,Wang1,Niset,Yuan,He24,Jiang,zhou2022orthogonal,zhou2023npartite,zhang2015nonlocality,croke2017difficulty,halder2019strong,li2022genuine}.

More recently, Halder \emph{et al.}~\cite{halder2019strong} introduced the notion of \emph{strong quantum nonlocality}.

Strongly nonlocal UPBs are locally irreducible across every bipartition, and, as UPBs, their orthogonal complements contain no product states \cite{bennett1999upb,zhang2020locally,shi2020unextendible,shi2022strongly,cohen2023local}. It is well known that suitable entanglement resources can enhance local distinguishability of orthogonal sets \cite{zhou2022orthogonal,zhang2020locally,bandyopadhyay2018optimal,cohen2007local,cohen2008understanding,rout2019genuinely,zhang2018local}. Consequently, optimizing the use of low-dimensional entanglement has become an active area of research \cite{zhang2020locally,cohen2008understanding,zhang2018local,yan2025variational}. In particular, Cohen~\cite{cohen2008understanding} identified UPBs that become LOCC-distinguishable with the aid of medium-dimensional maximally entangled states. Later, Zhang \emph{et al.}~\cite{zhang2018local} showed that multiple copies of \(2\otimes 2\) maximally entangled states (EPR pairs) can replace higher-dimensional entanglement, thereby reducing experimental complexity.

Entanglement is an exceptionally valuable quantum resource \cite{Horodecki09,gao2014permutationally,gao2010detection,Yan}. In practice, low-dimensional entangled states are appealing due to their relative ease of generation and control. Incorporating them into discrimination protocols can reduce both the entanglement cost and the operational overhead \cite{morelli2023resource,Du2024ErrorheraldedHQ}. Therefore, it is of considerable interest to develop efficient LOCC discrimination schemes relying only on low-dimensional entanglement.

In this work, we further investigate entanglement-assisted local discrimination of UPBs. We first introduce the preliminaries and relevant definitions. We then consider a known strongly nonlocal UPB in a \(3\otimes 3\otimes 3\) system \cite{shi2022strongly} and propose two classes of LOCC protocols: one class uses teleportation, while the other avoids teleportation by distributing low-dimensional entanglement more effectively and coordinating multipartite local measurements. Building on these ideas, we generalize the schemes to \(d\otimes d\otimes d\) systems, compare resource costs, and discuss the trade-offs among different protocols. We conclude with a brief summary.

\section{Preliminaries}
In this section, we introduce basic concepts and notation used throughout the paper.

\textbf{\emph{Definition~1} \cite{bennett1999upb}.}
A finite set of mutually orthogonal normalized product states
\(
\mathcal S
=
\left\{
\bigotimes_{j=1}^{m}
|\psi_j^{(i)}\rangle
\right\}_{i=1}^{n}
\subset
\bigotimes_{j=1}^{m}\mathbb C^{d_j}
\)
is an unextendible product basis if its elements are mutually
orthogonal product states,
\(
n<\prod_{j=1}^{m}d_j,
\)
and no nonzero product state is orthogonal to every element of
\(\mathcal S\). Equivalently, a UPB cannot be extended by adding any additional product state while preserving mutual orthogonality. Consequently, its orthogonal complement is a completely entangled subspace (CES).

\textbf{\emph{Definition~2} \cite{halder2019strong}.}
A set of orthogonal product states
\(\{\otimes_{j=1}^{m}|\psi\rangle_{j}^{i}\}_{i=1}^{n}\)
on \(\otimes_{j=1}^{m}\mathbb{C}^{d_{j}}\) is said to be \emph{strongly nonlocal} if it is locally irreducible across every bipartition; namely, for every bipartition, no party in any bipartition can initiate a nontrivial orthogonality-preserving measurement.

\textbf{\emph{Definition~3} \cite{zhou2022orthogonal}.}
In local state discrimination,  an \emph{entanglement resource configuration}
specifies the types, numbers, local dimensions, and sharing parties
of the entangled resource states supplied to the protocol.

We denote such a configuration by
\(
\left\{
(N,|\phi^+(d_1)\rangle_{P_1P_2});
(M,|\phi^+(d_2)\rangle_{P_2P_3})
\right\},
\)
where \(N\) and \(M\) are the numbers of supplied resource states and \(|\phi^+(d)\rangle=\frac{1}{\sqrt{d}}\sum_{i=0}^{d-1}|ii\rangle\) is a maximally entangled state.
The average entanglement consumption of a protocol is defined as the expected amount of entanglement resources that are actually consumed during the protocol.

\textbf{\emph{Definition~4}.}
For any real number \(x\), let \(\lceil x\rceil\) and
\(\lfloor x\rfloor\) denote the ceiling and floor functions,
respectively. For later use, define
\(
\ell=\left\lceil\frac{d}{2}\right\rceil,
\widetilde{\ell}=\left\lfloor\frac{d}{2}\right\rfloor,
\omega_n=\exp\left(\frac{2\pi\mathrm{i}}{n}\right). \) The quantities \(\ell\) and \(\widetilde{\ell}\) denote the upper and lower halves of the local dimension, respectively. We denote by
\(
\mathbb Z_n=\{0,1,\ldots,n-1\}
\)
the cyclic group of integers modulo \(n\). Note that
\(\ell+\widetilde{\ell}=d\).

Based on these definitions, we design local discrimination protocols for UPBs in \(d\otimes d\otimes d\) systems under various resource-allocation scenarios, and we quantify the corresponding entanglement consumption for each protocol.

\section{Entanglement-assisted discrimination in the \(3\otimes3\otimes3\) system}
We consider the following strongly nonlocal UPB in a \(3 \otimes 3 \otimes 3\) system \cite{shi2022strongly}:
\begin{equation}\label{333}
 \begin{aligned}
	\mathcal{A}_1 &:= \bigl\{ |\xi_j\rangle_A |0\rangle_B |\eta_i\rangle_C \mid (i,j) \in \left(\mathbb{Z}_2\times\mathbb{Z}_2\right)\setminus\{(0,0)\} \bigr\}, \\
	\mathcal{A}_2 &:= \bigl\{ |\xi_j\rangle_A |\eta_i\rangle_B |2\rangle_C \mid (i,j) \in \left(\mathbb{Z}_2\times\mathbb{Z}_2\right)\setminus\{(0,0)\} \bigr\}, \\
	\mathcal{A}_3 &:= \bigl\{ |2\rangle_A |\xi_j\rangle_B |\eta_i\rangle_C \mid (i,j) \in \left(\mathbb{Z}_2\times\mathbb{Z}_2\right)\setminus\{(0,0)\} \bigr\}, \\
	\mathcal{B}_1 &:= \bigl\{ |\eta_i\rangle_A |2\rangle_B |\xi_j\rangle_C \mid (i,j) \in \left(\mathbb{Z}_2\times\mathbb{Z}_2\right)\setminus\{(0,0)\} \bigr\}, \\
	\mathcal{B}_2 &:= \bigl\{ |\eta_i\rangle_A |\xi_j\rangle_B |0\rangle_C \mid (i,j) \in \left(\mathbb{Z}_2\times\mathbb{Z}_2\right)\setminus\{(0,0)\} \bigr\}, \\
	\mathcal{B}_3 &:= \bigl\{ |0\rangle_A |\eta_i\rangle_B |\xi_j\rangle_C \mid (i,j) \in \left(\mathbb{Z}_2\times\mathbb{Z}_2\right)\setminus\{(0,0)\} \bigr\}, \\
	|S\rangle &= \left( \sum_{i=0}^2 |i\rangle \right)_A \left( \sum_{j=0}^2 |j\rangle \right)_B \left( \sum_{k=0}^2 |k\rangle \right)_C.
\end{aligned}
\end{equation}
Here \(|\eta_i\rangle = |0\rangle + (-1)^i|1\rangle\) and \(|\xi_j\rangle = |1\rangle + (-1)^j|2\rangle\) for \(i,j \in \mathbb{Z}_2\). The vectors below are written without normalization factors, which do not affect orthogonality relations. The UPB
\(
\bigcup_{i=1}^3 \bigl( \mathcal{A}_i \cup \mathcal{B}_i \bigr) \cup \bigl\{ |S\rangle \bigr\}
\)
cannot be perfectly distinguished by LOCC. Because it is a low-dimensional example of strong nonlocality without
entanglement in the \(3\otimes3\otimes3\) setting, an important question is the following: \emph{How efficiently can entanglement assist the perfect local discrimination of this UPB?}
To address this, we propose three entanglement-assisted discrimination schemes. By exploiting the block structure of the set, our protocols reduce entanglement consumption compared with the naive approach of teleporting the entire multipartite state to a single party.

\textbf{\emph{Theorem 1}.}
The entanglement resource configuration
\(\left\{(1, |\phi^+(2)\rangle_{AB}); (1, |\phi^+(3)\rangle_{BC})\right\}\)
is sufficient for perfectly distinguishing the UPB~(\ref{333}) by LOCC.

\emph{Proof.}
Let the data system be shared by Alice, Bob, and Charlie. The labels $A,B,C$ refer exclusively to the original UPB subsystems. Using the shared ancillary maximally entangled state \(|\phi^+(3)\rangle_{bc}\), Charlie first teleports the data subsystem \(C\) to Bob. Let \(\widetilde{B}\) denote Bob's enlarged system comprising his original subsystem \(B\) and the teleported subsystem \(C\). Alice and Bob additionally share an EPR state
\(|\phi^+(2)\rangle_{ab} = \frac{1}{\sqrt{2}}(|00\rangle_{ab} + |11\rangle_{ab}\)).
The initial joint state can be written as
\begin{equation}\label{e3.2}
|\psi\rangle_{A\widetilde{B}} \otimes |\phi^+(2)\rangle_{ab}.
\end{equation}
For brevity, we write \(|ij\rangle = |i\rangle\otimes|j\rangle = |3i+j\rangle\). Since the subsets are mutually orthogonal and each subset can be perfectly discriminated after being identified, it is sufficient to construct an orthogonality-preserving protocol that identifies the subset labels.

\emph{Step~1.} Alice performs the measurement
\[
\mathcal{M}_1\equiv\{M_{11}:=P[(|0\rangle,|1\rangle)_A;|0\rangle_a]+P[|2\rangle_A;|1\rangle_a],
\;M_{12}:=I - M_{11}\}.
\]
where $P[X;Y]$ denotes the projector onto the tensor-product subspace generated by $X$ and $Y$.
Here, \(P\big[(|0\rangle, |1\rangle)_A; |0\rangle_a\big]
:= (|0\rangle\langle0|+|1\rangle\langle1|)_A\otimes(|0\rangle\langle0|)_a\) with analogous notation used throughout.

\begin{figure}[h]
	\centering
	\includegraphics[width=0.78\textwidth]{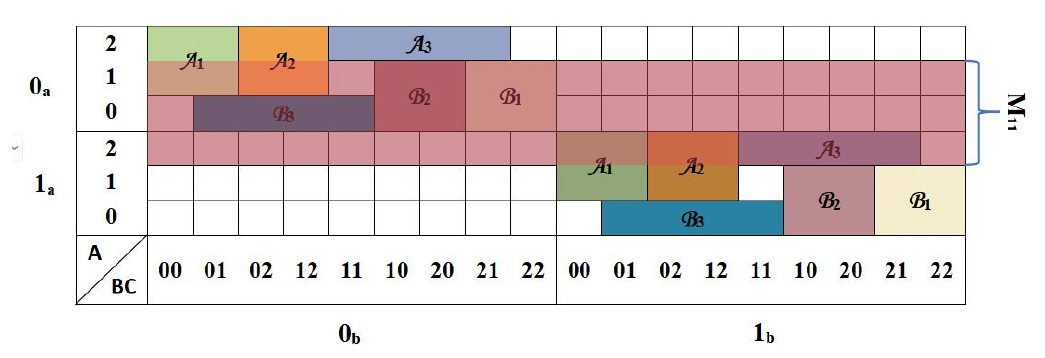}
	\caption{With the shared EPR state \( |\phi^+(2)\rangle_{ab}\), the state in Eq.~(\ref{e3.2}) can be represented on a \(2d_A\times 2d_Bd_C\) grid. The lavender-shaded region indicates the support of \(M_{11}\). \label{t3.1}}
\end{figure}

Conditioned on obtaining outcome \(M_{11}\) (cf. Fig.~\ref{t3.1}), the post-measurement states become
\begin{align*}
	&\mathcal{A}_1 \to \left\{ \left( |1\rangle_A |00\rangle_{ab} + (-1)^j |2\rangle_A |11\rangle_{ab} \right) |0\circ\eta_i\rangle_{\widetilde{B}} \right\}, \\
	&\mathcal{A}_2 \to \left\{ \left( |1\rangle_A |00\rangle_{ab} + (-1)^j |2\rangle_A |11\rangle_{ab} \right) |\eta_i\circ2\rangle_{\widetilde{B}} \right\}, \\
	&\mathcal{A}_3 \to \left\{  |2\rangle_A  |\xi_j\circ\eta_i\rangle_{\widetilde{B}}|11\rangle_{ab} \right\}, \\
	&\mathcal{B}_1 \to \left\{  |\eta_i\rangle_A  |2\circ\xi_j\rangle_{\widetilde{B}}|00\rangle_{ab} \right\}, \\
	&\mathcal{B}_2 \to \left\{  |\eta_i\rangle_A  |\xi_j\circ0\rangle_{\widetilde{B}}|00\rangle_{ab} \right\}, \\
	&\mathcal{B}_3 \to \left\{  |0\rangle_A  |\eta_i\circ\xi_j\rangle_{\widetilde{B}}|00\rangle_{ab} \right\}, \\
	&|S \rangle \to \left( \left( |0\rangle + |1\rangle \right)_A |00\rangle_{ab} + |2\rangle_A |11\rangle_{ab} \right) \left|\gamma \circ\gamma\right\rangle_{\widetilde{B}},
\end{align*}
where \(|\gamma\rangle=|0\rangle+|1\rangle+|2\rangle\).
Throughout the remainder of this paper, the symbol ``\(\circ\)'' denotes tensor concatenation of subsystems:
for states \(|\psi_1\rangle_B\) and \(|\psi_2\rangle_C\),
\(
|\psi_1 \circ \psi_2\rangle_{\widetilde{B}} = |\psi_1\rangle_B\otimes|\psi_2\rangle_C.
\)
In particular,
\(
\left| (0,\dots,d_B-1) \circ (0,\dots,d_C-1) \right\rangle_{\widetilde{B}}
\)
denotes the computational basis of \(\widetilde{B}\) ordered lexicographically.

$Step~2.$ Bob performs
		\[
		\begin{aligned}
		\mathcal{M}_2 \equiv \left\{ \right.
		M_{21}&:=P\left[ \left| (1,2) \circ (0,1) \right\rangle_{\widetilde{B}} ; |1\rangle_b \right], \\
		\quad \quad \quad \quad M_{22}&:= P\left[ \left| 2 \circ (1,2) \right\rangle_{\widetilde{B}} ; |0\rangle_b \right],\\
		\quad \quad \quad \quad M_{23}&:= P\left[  \left| (1,2) \circ 0 \right\rangle_{\widetilde{B}} ; |0\rangle_b \right],\\
		\quad \quad \quad \quad	M_{24}&:=I - M_{21}-M_{22}-M_{23} \left. \right\}.
	\end{aligned}
	\]
	\begin{figure}[h]
	\centering
	\includegraphics[width=0.75\textwidth]{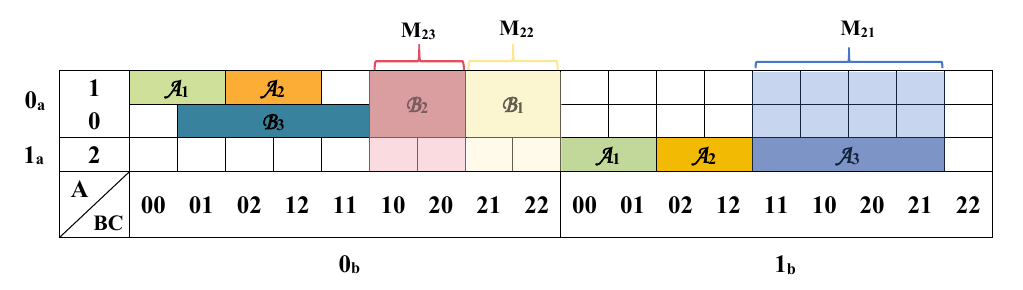}
	\caption{The lightly shaded regions represent the measurement outcomes in the \(M_{21}\), \(M_{22}\), and \(M_{23}\) directions for Bob, respectively. \label{t3.2}}
\end{figure}

The measurement outcomes corresponding to the operators \( M_{21} \), \( M_{22} \), \( M_{23} \)  (see Fig.~\ref{t3.2}) are
	\[
	\begin{aligned}
		M_{21} &\Rightarrow \mathcal{A}_{3} \text{ and } |S \rangle \to |2\rangle_A |11\rangle_{ab}		(|10\rangle+|11\rangle+|20\rangle+|21\rangle)_{\widetilde{B}},\\
		M_{22} &\Rightarrow \mathcal{B}_{1} \text{ and } |S \rangle \to (|0\rangle+|1\rangle)_A |00\rangle_{ab}	(|21\rangle+|22\rangle)_{\widetilde{B}},  \\
		M_{23} &\Rightarrow \mathcal{B}_{2} \text{ and } |S \rangle \to (|0\rangle+|1\rangle)_A |00\rangle_{ab}	(|10\rangle+|20\rangle)_{\widetilde{B}},
	\end{aligned}
	\]
respectively. Indeed, after the corresponding subset label is identified, the remaining states differ only on Bob's subsystem and can therefore be distinguished by Bob's local measurement. If $M_{24}$ clicks, the state falls into one of the remaining subsets $\mathcal{A}_1$, $\mathcal{A}_2$, $\mathcal{B}_3$ and  state \(|S \rangle \to \left( |0\rangle + |1\rangle \right)_A |00\rangle_{ab} (|00\rangle+|01\rangle+|02\rangle+|11\rangle+|12\rangle)_{\widetilde{B}}+|2\rangle _A |11\rangle_{ab} (|00\rangle+|01\rangle+|02\rangle+|12\rangle+|22\rangle)_{\widetilde{B}}.\)
	
\begin{figure}[h]
	\centering
	\includegraphics[width=0.58\textwidth]{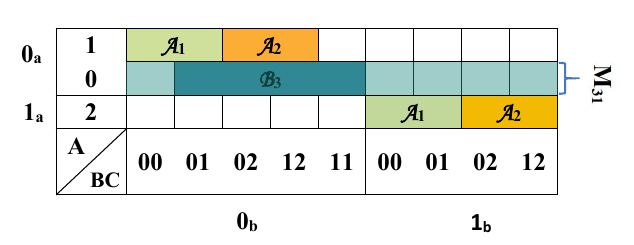}
	\caption{The measurement result for direction \(M_{31}\) in Alice's subsystem corresponds to the pale green region. \label{t3.3}}
\end{figure}

$Step~3.$ Alice performs
\[\mathcal{M}_3 \equiv \left\{ \right.
M_{31}:=P[|0\rangle_{A} ; |0\rangle_{a}],M_{32}:=I - M_{31} \left. \right\}.\]
As shown in Fig.~\ref{t3.3}, if $ M_{31}  $ clicks, the resulting subsets are \(\mathcal{B}_{3}\) and  \(\{|S\rangle \}\to \{|0\rangle_A |00\rangle_{ab}   (|00\rangle+|01\rangle+|02\rangle+|11\rangle+|12\rangle)_{\widetilde{B}}\}\), which  are locally distinguishable. Otherwise, the given subset is one of the following $\mathcal{A}_1,$ $ \mathcal{A}_2$ and  \(\bigl\{|S\rangle\bigr\} \to  \{|1\rangle_A |00\rangle_{ab} (|00\rangle+|01\rangle+|02\rangle+|11\rangle+|12\rangle)_{\widetilde{B}} +  |2\rangle_A |11\rangle_{ab} (|00\rangle+|01\rangle+|02\rangle+|12\rangle+|22\rangle)_{\widetilde{B}}\} .\)

$Step~4.$ Bob performs
\[\mathcal{M}_4 \equiv \left\{ \right.
M_{41}:=P[(|00\rangle,|01\rangle)_{\widetilde{B}}; (|0\rangle,|1\rangle)_{b}],M_{42}:=I - M_{41} \left. \right\}.\] If $ M_{41}$ clicks, the resulting subsets are \(\mathcal{A}_{1}\) and \(\{|S \rangle\} \to\{(|1\rangle_A |00\rangle_{ab} +|2\rangle_A |11\rangle_{ab}) (|00\rangle+|01\rangle)_{\widetilde{B}}\}\). Otherwise, the subsets are \( \mathcal{A}_2\) and  \(\{|S \rangle\}\to\{|1\rangle_A |00\rangle_{ab}  (|02\rangle+|11\rangle+|12\rangle)_{\widetilde{B}}+|2\rangle _A |11\rangle_{ab} (|02\rangle+|12\rangle+|22\rangle)_{\widetilde{B}}\}\). These subsets are therefore LOCC-distinguishable.

If \( M_{12} \) clicks in Step~1, the complementary branch can be treated by an analogous discrimination protocol. ~ \hfill $\square$
	
In every identified branch, the projected stopper state remains
locally distinguishable from the states in the corresponding subset.
For readability, the explicit stopper-state components will therefore
be omitted in the subsequent protocols.

\textbf{\emph{Theorem 2}.}
The resource configuration
\(\left\{(1, |\phi^+(2)\rangle_{ab}); (1, |\phi^+(2)\rangle_{ac})\right\}\)
is sufficient for perfectly distinguishing the UPB~(\ref{333}) by LOCC.

\emph{Proof.}
Assume that Alice--Bob and Alice--Charlie each share an EPR state.
Bob performs
\[
\mathcal{M}_1\equiv \{ M_{11}:=P[|0\rangle_B;|0\rangle_{b_1}]+P[(|1\rangle,|2\rangle)_B;|1\rangle_{b_1}] ,\; M_{12}:=I - M_{11}\},
\]
and Charlie performs
\[
\mathcal{M}_2\equiv \{ M_{21}:=P[(|0\rangle,|1\rangle)_C;|0\rangle_{c_1}]+P[|2\rangle_C;|1\rangle_{c_1}],\; M_{22}:=I - M_{21}\}.
\]
Since the corresponding measurement operators act on disjoint subsystems,
all operators from the two measurements commute. Bob and Charlie then
communicate their outcomes to Alice before the subsequent adaptive
measurements are performed. Conditioned on the outcomes \(M_{11}\) and \(M_{21}\), the resulting post-measurement states are
\[	\begin{aligned}
			&\mathcal{A}_1 \to  \bigl\{ |\xi_{j}\rangle_A|0\rangle_B|\eta_{i}\rangle_C|00\rangle_{a_1b_1}|00\rangle_{a_2c_1}  \bigr\}, \\
			&\mathcal{A}_2 \to  \bigl\{ |\xi_{j}\rangle_A(|0\rangle_B|00\rangle_{a_1b_1}+(-1)^{i}|1\rangle_B|11\rangle_{a_1b_1})|2\rangle_C|11\rangle_{a_2c_1}  \bigr\}, \\
			&\mathcal{A}_3 \to  \bigl\{ |2\rangle_A|\xi_j\rangle_B |11\rangle_{a_1b_1} |\eta_i\rangle_C |00\rangle_{a_2c_1} \bigr\}, \\
				&\mathcal{B}_1 \to  \bigl\{|\eta_i\rangle_A|2\rangle_B|11\rangle_{a_1b_1}(|1\rangle_C|00\rangle_{a_2c_1}+(-1)^{j}|2\rangle_C|11\rangle_{a_2c_1}) \bigr\}, \\
			&\mathcal{B}_2 \to  \bigl\{ |\eta_i\rangle_A|\xi_j\rangle_B|11\rangle_{a_1b_1}|0\rangle_C|00\rangle_{a_2c_1}    \bigr\},
		\end{aligned}\]
\[	\begin{aligned}
		&	\mathcal{B}_3 \to \bigl\{ |0\rangle_A (|0\rangle_B|00\rangle_{a_1b_1}+(-1)^{i}|1\rangle_B|11\rangle_{a_1b_1})(|1\rangle_C|00\rangle_{a_2c_1}+(-1)^{j}|2\rangle_C|11\rangle_{a_2c_1})  \bigr\}.
		\end{aligned}\]
The ancillary registers record the corresponding computational-sector information.

$Step~2.$ Alice performs the measurement
	\begin{align*}
		\mathcal{M}_3  &= \bigl\{
		M_{31}:=P[(|1\rangle,|2\rangle)_{A} ; |0\rangle_{a_1}; |0\rangle_{a_2}], M_{32}:=P[|2\rangle_{A} ; |1\rangle_{a_1}; |0\rangle_{a_2}], M_{33}:=I - M_{31}-M_{32}\bigr\}.
	\end{align*}
The outcomes corresponding to \(M_{31}\) and \(M_{32}\) identify
\(\mathcal A_1\) and \(\mathcal A_3\), respectively. Once the subset
label is known, the states within each subset can be distinguished by
appropriate local measurements performed by the relevant parties. The other outcome, \(M_{33}\), leaves \(\mathcal{A}_{2}\) and \(\mathcal{B}_{i}\) with \(i=1,2,3\).
	
$Step~3.$ Charlie performs the measurement\[\mathcal{M}_4 \equiv \left\{ \right.
	M_{41}:=P[|0\rangle_{C} ; |0\rangle_{c_1}],M_{42}:=I - M_{41}\left.  \right\}.\] If \( M_{41}\) clicks, we obtain the subset \(\mathcal{B}_{2}\) which is perfectly LOCC distinguishable.
	
$Step~4.$ Bob performs the measurement\[\mathcal{M}_5 \equiv \left\{ \right.
	M_{51}:=P[|2\rangle_{B} ; |1\rangle_{b_1}],M_{52}:=I - M_{51}\left.  \right\}.\] If \( M_{51}\) clicks, the subset is \(\mathcal{B}_{1}\) which is perfectly LOCC distinguishable. \( M_{52}\) is a projection operator acting in Bob's party, it leaves  \(\mathcal{A}_{2}\) and \(\mathcal{B}_{3}\).
	
$Step~5.$ Alice performs the measurement\[\mathcal{M}_6 \equiv \left\{ \right.
	M_{61}:=P[(|1\rangle,|2\rangle)_{A} ; (|0\rangle,|1\rangle)_{a_1}; |1\rangle_{a_2}],	M_{62}:=I - M_{61} \left. \right\}.\]
	If \( M_{61}\) clicks, the corresponding subset is \(\mathcal{A}_{2}\). Otherwise, the subset is \(\mathcal{B}_{3}\). These are perfectly LOCC distinguishable.	
	
	In addition, if other operators click in Step~1, we can find similar discrimination schemes. ~ \hfill $\square$

\textbf{\emph{Theorem 3}.}
The UPB in Eq.~\eqref{333} can be perfectly distinguished by LOCC
with one shared three-qutrit GHZ state
\[
|\mathrm{GHZ}_3\rangle_{abc}
=
\frac{1}{\sqrt3}\sum_{r=0}^{2}|rrr\rangle_{abc}.
\]

\emph{Proof.}
To distinguish the UPB locally~(\ref{333}), let Alice, Bob, and Charlie share the ancillary three-qutrit GHZ state
\(|\mathrm{GHZ}_3\rangle_{abc}\). Now, we only need to locally distinguish these subsets \(\mathcal{A}_{i}\) and \(\mathcal{B}_{i}\) for \(i=1,2,3\).
	
$Step~1.$ Alice performs a measurement
	\[\mathcal{M}_1\equiv \{ M_{11}:=P[|0\rangle_A;|0\rangle_{a}]+P[(|1\rangle,|2\rangle)_A;(|1\rangle,|2\rangle)_{a}] , M_{12}:=I - M_{11}\}.\]
	Charlie performs a measurement
	\[\mathcal{M}_2\equiv \{ M_{21}:=P[(|0\rangle,|1\rangle)_C;|1\rangle_{c}]+P[|2\rangle_C;|2\rangle_{c}]+P[I_C;|0\rangle_{c}] , M_{22}:=I - M_{21}\}.\]
Conditioned on outcomes \(M_{11}\) and \(M_{21}\), the post-measurement states are as follows.
\begin{align*}
		&\mathcal{A}_1 \to  \bigl\{  |\xi_{j}\rangle_A|0\rangle_B |\eta_{i}\rangle_{C} |111\rangle_{abc} \bigr\}, \\
		&\mathcal{A}_2 \to  \bigl\{  |\xi_{j}\rangle_A|\eta_{i}\rangle_B |2\rangle_{C} |222\rangle_{abc}\bigr\}, \\
		&\mathcal{A}_3 \to  \bigl\{|2\rangle_A|\xi_{j}\rangle_B |\eta_{i}\rangle_{C} |111\rangle_{abc}\bigr\}, \\
		&\mathcal{B}_1 \to  \bigl\{|0\rangle_A|2\rangle_B |\xi_{j}\rangle_{C} |000\rangle_{abc}+(-1)^{i}|1\rangle_A|2\rangle_B(|1\rangle_{C} |111\rangle_{abc}+(-1)^{j}|2\rangle_{C} |222\rangle_{abc}) \bigr\}, \\
		&\mathcal{B}_2 \to  \bigl\{ (|0\rangle_{A}|000\rangle_{abc}+(-1)^{i}|1\rangle_{A}|111\rangle_{abc} )|\xi_{j}\rangle_{B}|0\rangle_{C} \bigr\}, \\
		&\mathcal{B}_3 \to \bigl\{|0\rangle_A|\eta_{i}\rangle_B |\xi_{j}\rangle_{C} |000\rangle_{abc} \bigr\}.
	\end{align*}
	
$Step~2.$ Bob performs the measurement\[\mathcal{M}_3 \equiv \left\{ \right.
	M_{31}:=P[|0\rangle_{B} ; |1\rangle_{b}],M_{32}:=P[(|0\rangle,|1\rangle)_{B} ; |2\rangle_{b}],	M_{33}:=I - M_{31}-M_{32} \left. \right\}.\]
The measurement outcomes corresponding to \(M_{31}\) and \(M_{32}\) are \(\mathcal{A}_{1}\) and \(\mathcal{A}_{2}\), respectively. For operator \(M_{33}\), the surviving candidates belong to \(\mathcal A_3\cup\mathcal B_1\cup\mathcal B_2\cup\mathcal B_3.\)

$Step~3.$ Alice performs the measurement\[\mathcal M_4 \equiv \left\{ \right. M_{41}:=P[|2\rangle_{A} ; |1\rangle_{a}],M_{42}:=I - M_{41}\left.  \right\}.\] If \( M_{41}\) clicks, the subset is \(\mathcal{A}_{3}\). Otherwise, the given state belongs to one of the remaining subsets \(\mathcal{B}_{1}\), \(\mathcal{B}_{2}\), and \(\mathcal{B}_{3}\).
	
$Step~4.$ Charlie performs the measurement\[\mathcal{M}_5 \equiv \left\{ \right.
	M_{51}:=P[|0\rangle_{C} ;( |0\rangle,|1\rangle)_{c}],M_{52}:=I - M_{51}\left.  \right\}.\]
	The outcome corresponding to the measurement \(M_{51}\) is \(\mathcal{B}_{2}\). Otherwise, the state belongs to one of the subsets \(\mathcal{B}_{1}\) and \(\mathcal{B}_{3}\).
	
$Step~5.$ Bob performs the measurement\[\mathcal{M}_6 \equiv \left\{ \right.
	M_{61}:=P[(|0\rangle,|1\rangle)_{B} ;|0\rangle_{b}],M_{62}:=I - M_{61}\left.  \right\}.\]
	The outcome corresponding to the measurement $M_{61}$ is $\mathcal{B}_{3}$. Otherwise, the state is an element of the subset \(\mathcal{B}_{1}\). If another operator clicks in Step~1, we can obtain a similar entanglement-assisted discrimination protocol to distinguish set (\ref{333}).
\hfill $\square$

\medskip
\noindent

\textbf{Resource comparison.}
The resource cost of Theorem~1 is \(1+\log_2 3\) ebits, including the qutrit entanglement consumed for teleportation. Theorem~2 avoids teleportation and achieves the same task using two EPR pairs, i.e., \(2\) ebits. Theorem~3 also avoids teleportation and employs a single genuinely tripartite entangled GHZ state, namely \( |\mathrm{GHZ}_3\rangle_{abc}\). Since this protocol relies on a genuinely tripartite entanglement resource, it is not assigned a numerical ebit cost under the present bipartite accounting convention. Instead, it provides an alternative multipartite resource that is sufficient for accomplishing the same discrimination task. In contrast, a strategy based solely on teleportation of full subsystems would require at least \(2\log_2 3\) ebits. The bipartite protocols require fewer ebits than the corresponding full-teleportation strategy. Furthermore, Theorem~2 shows that two EPR pairs provide a bipartite entanglement-assisted protocol achieving the same discrimination task, whereas the GHZ protocol uses a genuinely multipartite resource. Since EPR pairs are generally more accessible experimentally, the choice between the protocols of Theorems~2 and~3 can be made according to the available implementation constraints. We next generalize these results to \(d \otimes d \otimes d\) systems with \(d\ge 3\).

\section{Quantum state discrimination in \(d\otimes d\otimes d\)}
We recall the construction of strongly nonlocal UPB in \(d \otimes d\otimes d\) systems (\(d\ge 3\)) from Ref.~\cite{shi2022strongly}.

\begin{figure}[h]
	\centering
	\includegraphics[width=0.36\textwidth]{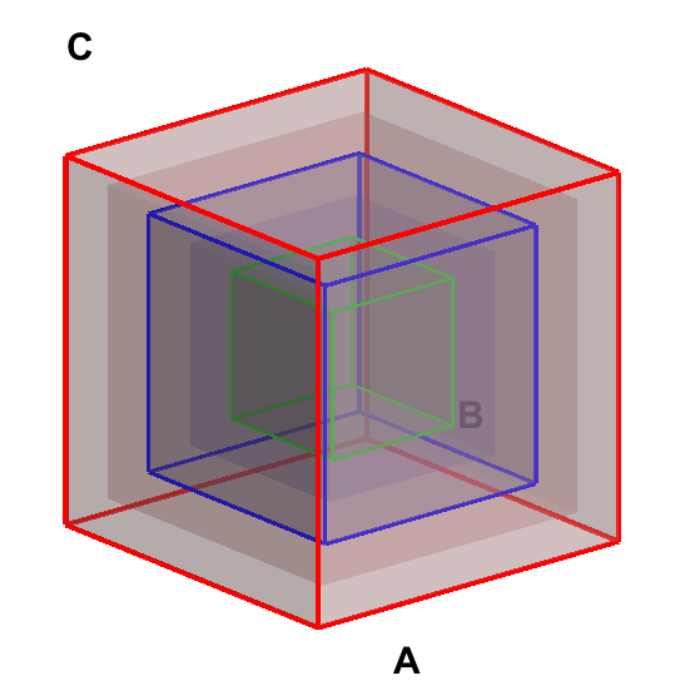}
	\caption{Layered cubic structure of the set given by Eq.~(\ref{ud}) in a \(d\otimes d\otimes d\) tripartite system. The outermost layer corresponds to \(k=0\) and the next layer to \(k=1\), and so forth. \label{jiegou}}
\end{figure}

\begin{equation}\label{e4.1}
\begin{aligned}
	&\mathcal{A}_1^{(d,d-2k)} := \left\{ \left| \xi_j^{(d-2k)} \right\rangle_A \left| k \right\rangle_B \left| \eta_i^{(d-2k)} \right\rangle_C \ \middle| \ (i,j) \in \mathbb{Z}_{d-1-2k} \times \mathbb{Z}_{d-1-2k} \setminus \{(0,0)\} \right\}, \\
	&\mathcal{A}_2^{(d,d-2k)} := \left\{ \left| \xi_j^{(d-2k)} \right\rangle_A \left| \eta_i^{(d-2k)} \right\rangle_B \left| d-1-k \right\rangle_C \ \middle| \ (i,j) \in \mathbb{Z}_{d-1-2k} \times \mathbb{Z}_{d-1-2k} \setminus \{(0,0)\} \right\}, \\
	&\mathcal{A}_3^{(d,d-2k)} := \left\{ \left| d-1-k \right\rangle_A \left| \xi_j^{(d-2k)} \right\rangle_B \left| \eta_i^{(d-2k)} \right\rangle_C \ \middle| \ (i,j) \in \mathbb{Z}_{d-1-2k} \times \mathbb{Z}_{d-1-2k} \setminus \{(0,0)\} \right\}, \\
	&\mathcal{B}_1^{(d,d-2k)} := \left\{ \left| \eta_i^{(d-2k)} \right\rangle_A \left| d-1-k \right\rangle_B \left| \xi_j^{(d-2k)} \right\rangle_C \ \middle| \ (i,j) \in \mathbb{Z}_{d-1-2k} \times \mathbb{Z}_{d-1-2k} \setminus \{(0,0)\} \right\}, \\
	&\mathcal{B}_2^{(d,d-2k)} := \left\{ \left| \eta_i^{(d-2k)} \right\rangle_A \left| \xi_j^{(d-2k)} \right\rangle_B \left| k \right\rangle_C \ \middle| \ (i,j) \in \mathbb{Z}_{d-1-2k} \times \mathbb{Z}_{d-1-2k} \setminus \{(0,0)\} \right\}, \\
	&\mathcal{B}_3^{(d,d-2k)} := \left\{ \left| k \right\rangle_A \left| \eta_i^{(d-2k)} \right\rangle_B \left| \xi_j^{(d-2k)} \right\rangle_C \ \middle| \ (i,j) \in \mathbb{Z}_{d-1-2k} \times \mathbb{Z}_{d-1-2k} \setminus \{(0,0)\} \right\},\\
	&\vert S_d\rangle := \left( \sum_{i=0}^{d-1} \vert i \rangle \right)_A \left( \sum_{j=0}^{d-1} \vert j \rangle \right)_B \left( \sum_{k=0}^{d-1} \vert k \rangle \right)_C,
\end{aligned}
\end{equation}
where
\(
|\eta_i^{(d - 2k)}\rangle = \sum_{t = k}^{d - 2 - k} \omega_{d - 1 - 2k}^{i(t-k)}|t\rangle
\)
and
\(
|\xi_j^{(d - 2k)}\rangle = \sum_{t = k}^{d - 2 - k} \omega_{d - 1 - 2k}^{j(t-k)}|t + 1\rangle
\)
for \(i,j \in \mathbb{Z}_{d - 1 - 2k}\) and \(k=0,\ldots,\ell-2\).
Let
\(
\mathcal{A}^{(d,0)}=\left\{ \vert \phi_r \rangle_A \vert \phi_s \rangle_B \vert \phi_t \rangle_C \mid (r,s,t)\in \mathbb{Z}_2^3\setminus\{(0,0,0)\}\right\}
\),
where
\(
\vert \phi_i \rangle = \left\vert \frac{d - 2}{2} \right\rangle + (-1)^i \left\vert \frac{d}{2} \right\rangle
\)
for \(i\in\mathbb{Z}_2\) (defined only when \(d\) is even). These sets form a strongly nonlocal UPB:
\begin{equation}\label{ud}
\begin{aligned}
\mathcal{U}_{d}=\begin{cases}
\displaystyle \bigcup_{k=0}^{\frac{d-3}{2}}\left[\bigcup_{i=1}^{3}
\left( \mathcal{A}_i^{(d,d-2k)}\cup \mathcal{B}_i^{(d,d-2k)} \right)\right]\cup \{ |S_d\rangle \}, & d~\text{odd},\\[0.6em]
\displaystyle \bigcup_{k=0}^{\frac{d-4}{2}}\left[\bigcup_{i=1}^{3}
\left( \mathcal{A}_i^{(d,d-2k)}\cup \mathcal{B}_i^{(d,d-2k)} \right)\right]\cup \{ |S_d\rangle \}\cup \mathcal{A}^{(d,0)}, & d~\text{even}.
\end{cases}
\end{aligned}
\end{equation}

Geometrically, the set $\mathcal{U}_{d}$ can be visualized as occupying a three-dimensional $d\times d\times d$ computational-basis lattice with a nested cubic-shell structure. As illustrated in Fig.~\ref{jiegou}, the shells are indexed by $k$, where $k=0$ denotes the outermost shell, $k=1$ the next inner shell, and increasing values of $k$ correspond to shells located progressively deeper inside the lattice. Apart from the central set \(\mathcal A^{(d,0)}\) for even \(d\), the non-stopper states in
\(\mathcal U_d\) are organized into nested cubic shells indexed by
\(k\). The computational-basis supports associated with different
noncentral shells are mutually disjoint. Consequently, every
non-stopper shell state has a unique shell index. The stopper state is
carried through the protocol and is separated together with the
corresponding identified branch. Based on this structural property, we propose three discrimination protocols.

\textbf{\emph{Theorem 4}.}
In \(d\otimes d\otimes d\), the strongly nonlocal UPB \(\mathcal{U}_{d}\) in Eq.~(\ref{ud}) can be perfectly distinguished by LOCC using the resource configuration
\(\left\{
(1,|\phi^+(\ell)\rangle_{ab});
(1,|\phi^+(d)\rangle_{bc})
\right\}\).

$Proof.$ Once a subset label has been identified, the states within each subset
in Eq.~\eqref{e4.1}, as well as those in
\(\mathcal A^{(d,0)}\) for even \(d\), can be perfectly distinguished
by LOCC. The protocol consists of three main stages. First, Charlie teleports subsystem \(C\) to Bob using the shared
maximally entangled state
 \(
|\phi^+(d)\rangle_{bc}
=
\frac{1}{\sqrt d}\sum_{i=0}^{d-1}|ii\rangle_{bc}\)
such that the quantum state is jointly owned by Alice and Bob and denoted as \(\lvert \psi \rangle_{A\widetilde{B}}\). Then, Alice and Bob share the maximally entangled state \( |\phi^+(\ell)\rangle_{ab}\) and obtain the initial state $|\psi\rangle_{A\widetilde{B}}\otimes|\phi^+(\ell)\rangle_{ab}$. Finally, a deterministic local discrimination protocol is designed based on this initial state. The proof is constructive and the complete sequence of local measurements is given in Appendix \ref{A}. ~ \hfill $\square$

With the results established above, the total supplied bipartite entanglement is
\(
\log_2d+\log_2\ell
=
\log_2(d\ell)
\)
ebits. The local dimension of $|\phi^+(\ell)\rangle$ is relatively high. Next, we provide a new resource configuration by replacing high-dimensional maximally entangled states \(|\phi^+(\ell
)\rangle\) with multiple copies of EPR states.

\textbf{\emph{Theorem~5}.}
In \(d\otimes d\otimes d\), the UPB \(\mathcal U_d\) in
Eq.~\eqref{ud} can be perfectly distinguished by LOCC using the
supplied resource configuration
\[
\left\{
(1,|\phi^+(d)\rangle_{bc});
(e,|\phi^+(2)\rangle_{ab})
\right\},
\]
where $e=\sum_{m=0}^{\ell-2}
\frac{(d-2m)^3-8(\ell-1-m)} {d^3-8(\ell-1)}.$

\emph{Proof.}
To realize the local discrimination of the quantum states in Eq.~(\ref{ud}), we proceed shell by shell according to the nested-shell structure of $\mathcal U_d$. Charlie first uses the entangled resource $|\phi^+(d)\rangle_{bc}$ to teleport subsystem $C$ to Bob, and the resulting composite subsystem is denoted by $\widetilde{B}$.

In the first round of the protocol, the total number of states is
\[
N_{\mathrm{tot}}=N_0=d^3-8(\ell-1).
\]
Alice and Bob share the first EPR pair
\[
|\phi^+(2)\rangle_{a_0b_0}
=\frac{1}{\sqrt2}
\left(|00\rangle+|11\rangle\right)_{a_0b_0}.
\]
Following the same procedure as in Theorem~1, they distinguish the subsets
$\mathcal{A}^{(d,d)}_{3}$,
$\mathcal{B}^{(d,d)}_{1}$,
$\mathcal{B}^{(d,d)}_{2}$,
$\mathcal{B}^{(d,d)}_{3}$,
$\mathcal{A}^{(d,d)}_{1}$, and
$\mathcal{A}^{(d,d)}_{2}$,
which constitute the outermost shell ($k=0$) of $\mathcal U_d$. The average entanglement consumption in this round is 1 ebit.

After the first round of measurements, the number of the remaining states is
\[
N_1=(d-2)^3-8(\ell-2).
\]
Alice and Bob share the second EPR pair
\[
|\phi^+(2)\rangle_{a_1b_1}
=\frac{1}{\sqrt2}
\left(|00\rangle+|11\rangle\right)_{a_1b_1},
\]
and similarly distinguish the subsets
$\mathcal{A}^{(d,d-2)}_{3}$,
$\mathcal{B}^{(d,d-2)}_{1}$,
$\mathcal{B}^{(d,d-2)}_{2}$,
$\mathcal{B}^{(d,d-2)}_{3}$,
$\mathcal{A}^{(d,d-2)}_{1}$, and
$\mathcal{A}^{(d,d-2)}_{2}$,
which belong to the shell $k=1$. Since the unknown state is chosen randomly from the set \(\mathcal U_d\), The average entanglement consumption in this round is $N_1/N_{\mathrm{tot}}$ ebits.

Repeating this procedure successively for $k=2,\ldots,\ell-2$ identifies all remaining shells. The average number of consumed EPR pairs is
\[
e
=
\sum_{m=0}^{\ell-2}\frac{N_m}{N_{\mathrm{tot}}}
=
\sum_{m=0}^{\ell-2}
\frac{(d-2m)^3-8(\ell-1-m)}
     {d^3-8(\ell-1)}.
\]
Accordingly, the average entanglement consumption of the protocol is
\[
\log_2 d+e
\]
ebits. Thus, perfect discrimination of this set requires only $\log_2 d + e$ ebits of entanglement on average. Details are provided in Appendix~\ref{B}.
\hfill $\square$

As shown in Fig.~\ref{fig:entanglement_vs_dimension}, Theorems~4 and~5 consume the same amount of entanglement when $d=4$. Theorem~5 is more efficient for $2<d/2\leq 12$, whereas Theorem~4 becomes more resource-efficient when $d/2>12$. We further observe that the effect of $(\ell-1)$ copies of EPR states are equivalent to \(|\phi^+(\ell)\rangle_{ab}\). We can replace a \(|\phi^+(\ell)\rangle_{ab}\) with multiple EPR states. From the perspective of experimental implementation, the multiple EPR states employed in Theorem 5 are easier to prepare than the single high-dimensional entangled state required in Theorem 4, making Theorem 5 simpler to implement experimentally. So, we can choose the scheme of Theorem 5 when the local dimension $d$ is small.

\textbf{\emph{Theorem 6}.} In \(d\otimes d\otimes d\), the resource configuration
\(\left\{ (1, |\phi^+(\ell)\rangle_{ab}); (1, |\phi^+(\ell)\rangle_{ac}) \right\}\)
is sufficient for perfect LOCC discrimination of \(\mathcal{U}_{d}\) in Eq.~(\ref{ud}).

$Proof.$ To achieve local distinguishability of the sets in Eq.~(\ref{ud}), let Alice share one Schmidt-rank-\(\ell\) maximally entangled state
with Bob and another independent Schmidt-rank-\(\ell\) maximally entangled state with Charlie. Then, the associated initial state is
\(|\psi\rangle_{ABC}
\otimes
|\phi^+(\ell)\rangle_{a_1b_1}
\otimes
|\phi^+(\ell)\rangle_{a_2c_1}.\)

After initial local measurements that encode the computational supports into the ancillary entangled registers by Bob and Charlie, the discrimination protocol proceeds through four additional measurement stages. First, Alice performs a local measurement to effectively distinguish the subsets $\mathcal{A}_1^{(d, d-2k)}$ and $\mathcal{A}_3^{(d, d-2k)}$. Second, we perform a measurement on party $C$ and identify the corresponding subset $\mathcal{B}_2^{(d,d-2k)}$. Next, Bob performs a local measurement that identifies
\(\mathcal B_1^{(d,d-2k)}\). Finally, we distinguish the remaining subsets based on the parity of $d$. If $d$ is odd, we only need to discriminate the subsets $\mathcal{A}_2^{(d, d-2k)}$ and $\mathcal{B}_3^{(d, d-2k)}$ by the local measurement on party $A$. If $d$ is even, the remaining possibilities
\(\mathcal A_2^{(d,d-2k)}\),
\(\mathcal B_3^{(d,d-2k)}\), and
\(\mathcal A^{(d,0)}\)
must be separated. Details are given in Appendix \ref{C}.~ \hfill $\square$

\begin{figure}[h]
	\centering
	\includegraphics[width=0.58\textwidth]{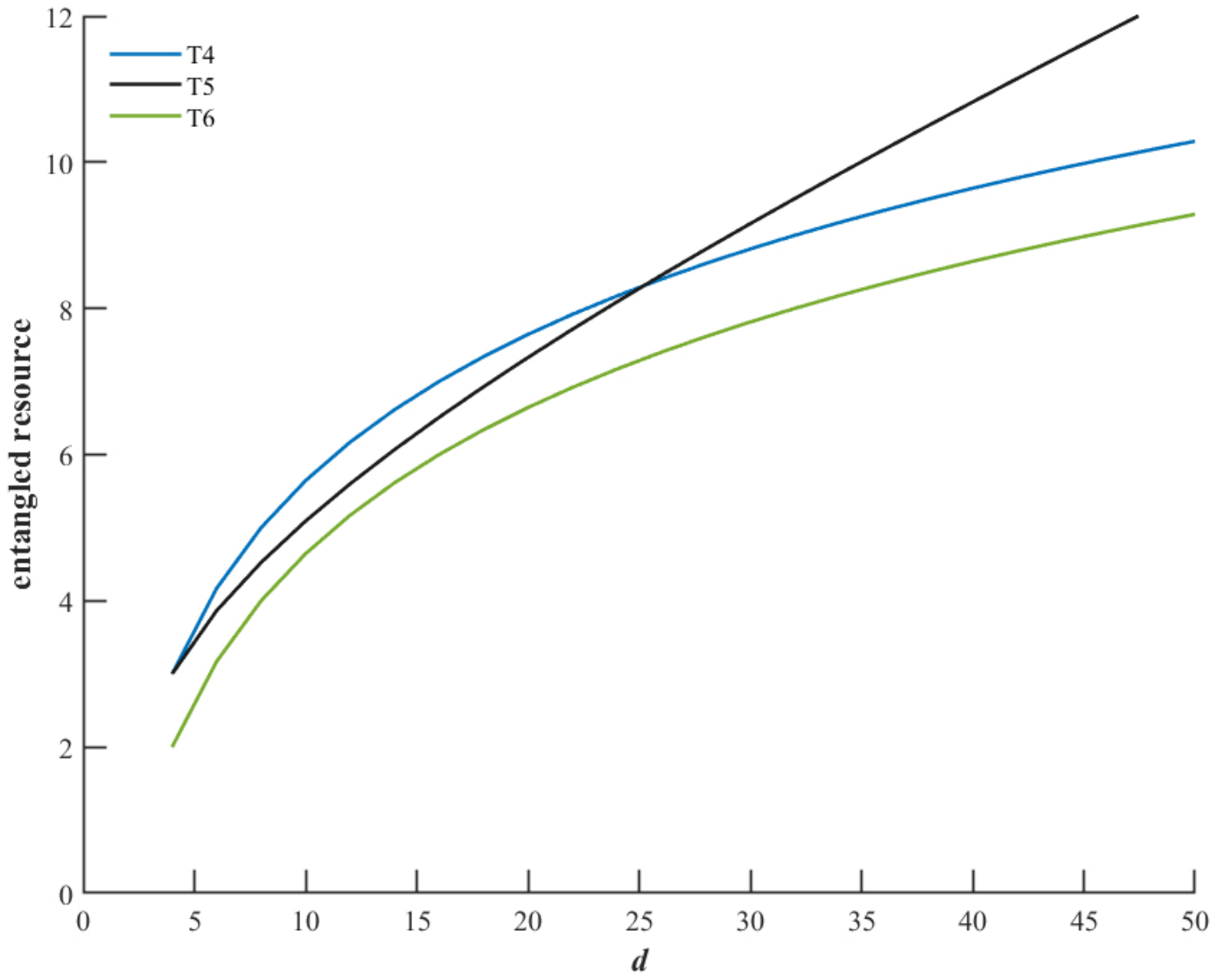}
	\caption{Comparison of the average amount of bipartite entanglement consumed. The curves for T4, T5, and T6 are the entanglement resources consumed on average in Theorems 4, 5, and 6, corresponding to even dimension $d$, respectively.}\label{fig:entanglement_vs_dimension}
\end{figure}

In this protocol, we do not use quantum teleportation and require only two \( |\phi^+(\ell )\rangle\) maximally entangled states. The total supplied entanglement is \( 2\log_2\ell \ \) ebits, which is smaller than the \(\log_2(d\ell)\)-ebit supplied cost of Theorem~4. Combined with Fig.~\ref{fig:entanglement_vs_dimension}, we observe that Theorems 4 and 5 consume more entanglement resources than Theorem 6 for dimensions $d \geq 3$, indicating that Theorem 6 is more resource-efficient under the adopted bipartite entanglement accounting convention. Specifically, among the evaluated protocols, Theorem 6 requires the lowest amount of supplied entanglement resources. Both Theorem 4 and Theorem 5 rely on one round of quantum teleportation. Overall, the teleportation-free protocol of Theorem~6 achieves the lowest supplied bipartite entanglement under the present resource accounting convention.

\section{Conclusion}
We studied entanglement-assisted LOCC discrimination of the strongly nonlocal UPBs introduced in Ref.~\cite{shi2022strongly}.
For the \(3\otimes3\otimes3\) example, we constructed a teleportation-free protocol using two EPR pairs. This result provides a partial answer to the problem raised in
Ref.~\cite{zhang2018local} concerning orthogonal-state families that can be deterministically distinguished by LOCC using several \(2\otimes2\) maximally entangled pairs without teleportation.

We then extended the constructions to \(d\otimes d\otimes d\) systems. Theorem~4 uses one Schmidt-rank-\(d\) pair for teleportation and one Schmidt-rank-\(\ell\) pair, whereas Theorem~5 replaces the latter resource by EPR pairs used shell by shell. Theorem~6 avoids teleportation and uses two Schmidt-rank-\(\ell\) maximally entangled pairs. Under the resource-accounting convention specified in the paper, Theorem~6 has the lowest bipartite entanglement cost among these three protocols for \(d\ge3\).

These results suggest that the allocation and dimensionality of shared entanglement can significantly influence the resource requirements of LOCC discrimination protocols. They also motivate further work on lower bounds and optimality: the present protocols establish sufficiency, but do not by themselves prove that the stated resources are minimal.
\section*{Funding}
This work was supported by the National Natural Science Foundation of China (Grant No.~12526564), the Hebei Natural Science Foundation (Grant No.~A2025403008), and the National Pre-research Funds of Hebei GEO University (Grant No.~KY2025YB15).

\begin{appendix}

%================================================
% Common notation (Appendix-wide conventions)
%================================================
\paragraph*{Appendix-wide notation and conventions.}
Throughout this appendix, Charlie teleports subsystem $C$ to Bob (when required), such that Bob effectively holds the joint subsystem \(\widetilde{B}:=B\otimes C .\) We use the symbol ``$\circ$'' as a shorthand notation for tensor products between subsystems $B$ and $C$, i.e.,
$|\psi_1\circ\psi_2\rangle_{\widetilde{B}}:=|\psi_1\rangle_B\otimes|\psi_2\rangle_C$.
For registers \(X_1,\ldots,X_q\), define
\[
P[\mathcal I_1{}_{X_1};\ldots;\mathcal I_q{}_{X_q}]
:=
\bigotimes_{\nu=1}^{q}
\left(
\sum_{i\in\mathcal I_\nu}
|i\rangle\langle i|_{X_\nu}
\right),
\]
where each \(\mathcal I_\nu\) is a set of computational-basis labels.
%================================================
\section{Proof of Theorem~4}\label{A}
%================================================
\paragraph*{Initial resources and state.}
Charlie and Bob initially share a Schmidt-rank-$d$ maximally entangled state $|\phi^+(d)\rangle_{bc}$ and perform one round of quantum teleportation so that Bob receives $C$. Hence, the unknown state is transformed into $|\psi\rangle_{A\widetilde{B}}$ with $\widetilde{B}:=B\otimes C$.
In addition, Alice and Bob share one maximally entangled state
\[
|\phi^+(\ell)\rangle_{ab}
=\frac{1}{\sqrt{\ell }}\sum_{r=0}^{\ell -1}|rr\rangle_{ab}.
\]
Therefore, the overall initial state for the discrimination stage is
$|\psi\rangle_{A\widetilde{B}}\otimes|\phi^+(\ell)\rangle_{ab}$, where $|\psi\rangle_{A\widetilde{B}}$ is any element of $\mathcal{U}_d$ in Eq.~\eqref{ud}.

\paragraph*{Step~1 (Alice).}
Alice performs the POVM
\[
\mathcal{M}_{1}\equiv \bigl\{M_{11},M_{12},\ldots,M_{1\ell}\bigr\},
\]
whose elements are
\[
\begin{aligned}
\mathcal{M}_{1} \equiv \biggl\{
	& M_{11} := P\left[\left(\lvert0\rangle,\, \lvert1\rangle,\, \dots,\, \lvert\widetilde{\ell}\rangle\right)_A;\, \lvert0\rangle_{a}\right]
	+ P\left[\lvert\widetilde{\ell} + 1\rangle_A;\, \lvert1\rangle_{a}\right] + \cdots \\
	&\quad\quad\quad + P\left[\lvert d-2\rangle_A;\, \lvert\ell - 2\rangle_{a}\right]
	+ P\left[\lvert d-1\rangle_A;\, \lvert\ell - 1\rangle_{a}\right], \\
	& M_{12} := P\left[\left(\lvert0\rangle,\, \lvert1\rangle,\, \dots,\, \lvert\widetilde{\ell} \rangle\right)_A;\, \lvert1\rangle_{a}\right]
	+ P\left[\lvert\widetilde{\ell} + 1\rangle_A;\, \lvert2\rangle_{a}\right] + \cdots \\
	&\quad \quad \quad + P\left[\lvert d-2\rangle_A;\, \lvert\ell  - 1\rangle_{a}\right]
	+ P\left[\, \lvert d-1\rangle_A;\, \lvert 0\rangle_{a}\right],\\
&\quad\quad\quad \vdots \\
	& M_{1\ell} := P\left[\left(\lvert0\rangle,\, \lvert1\rangle,\, \dots,\, \lvert\widetilde{\ell}\rangle\right)_A;\, \lvert\ell -1\rangle_{a}\right]
	+ P\left[\lvert\widetilde{\ell} + 1\rangle_A;\, \lvert0\rangle_{a}\right] + \cdots \\
	&\quad + P\left[\lvert d-2\rangle_A;\, \lvert\ell - 3\rangle_{a}\right]
	+ P\left[\lvert d-1\rangle_A;\, \lvert\ell -2\rangle_{a}\right]
\biggr\}.
\end{aligned}
\]
It is straightforward to verify that these operators are mutually orthogonal projectors satisfying
\(\sum_{m=1}^{\ell}M_{1m}=I_{Aa},\) thus defining a valid projective measurement. We analyze the branch corresponding to $M_{11}$; all other outcomes $M_{1m}$ ($m=2,\ldots,\ell $) are treated analogously and lead to the same conclusion.

Conditioned on $M_{11}$, the subsets transform as follows:
\[\begin{aligned}
	&\mathcal{A}_1^{(d, d-2k)} \to \Bigg\{ \Bigg[ \sum_{t=k}^{\widetilde{\ell}-1} \omega_{d-1-2k}^{j(t-k)} |t+1\rangle_A |00\rangle_{ab} + \omega_{d-1-2k}^{j(\widetilde{\ell}-k )} |\widetilde{\ell}+1 \rangle_A |11\rangle_{ab} + \cdots   \\
	& \quad\quad\quad \quad \quad\quad + \omega_{d-1-2k}^{j(d-2-2k)} |d-1-k\rangle_A  \left|\left(\ell-1-k\right) \left(\ell-1-k\right)\right\rangle _{ab} \Bigg] \left| k \circ \eta_i^{(d-2k)} \right\rangle_{\widetilde{B}} \Bigg\},\\
	&\mathcal{A}_2^{(d, d-2k)} \to \Bigg\{ \Bigg[ \sum_{t=k}^{\widetilde{\ell}-1} \omega_{d-1-2k}^{j(t-k)} |t+1\rangle_A |00\rangle_{ab}
	+ \omega_{d-1-2k}^{j( \widetilde{\ell}-k )} | \widetilde{\ell}+1 \rangle_A |11\rangle_{ab}
	+ \cdots   \\
	& \quad\quad\quad \quad\quad \quad  +\omega_{d-1-2k}^{j(d-2-2k)} \lvert d-1-k \rangle_A \left\lvert \left(\ell-1-k\right) \left(\ell-1-k \right)\right\rangle_{ab} \Bigg] \left| \eta_i^{(d-2k)} \circ (d-1-k) \right\rangle_{\widetilde{B}} \Bigg\},\\
	&\mathcal{A}_3^{(d, d-2k)} \to \Bigg\{ |d-1-k\rangle_A \left|(\ell -1-k)(\ell -1-k) \right\rangle _{ab} \left| \xi_j^{(d-2k)} \circ \eta_i^{(d-2k)} \right\rangle_{\widetilde{B}} \Bigg\}, \\
	&\mathcal{B}_1^{(d, d-2k)} \to \Bigg\{ \Bigg[ \sum_{t=k}^{\widetilde{\ell}} \omega_{d-1-2k}^{i(t-k)} |t\rangle_A |00\rangle_{ab} + \omega_{d-1-2k}^{i(\widetilde{\ell}+1-k)} | \widetilde{\ell}+1 \rangle_A |11\rangle_{ab} + \cdots \\
	&\quad \quad \quad \quad \quad \quad +\omega_{d-1-2k}^{i(d-2-2k)} |d-2-k\rangle_A
	\left| (\ell -2-k)(\ell -2-k) \right\rangle_{ab}\Bigg] \Bigl| (d-1-k) \circ \xi_j^{(d-2k)} \Bigr\rangle_{\widetilde{B}} \Bigg\},\\
\end{aligned}\]
\[\begin{aligned}
	&\mathcal{B}_2^{(d, d-2k)} \to \Bigg\{ \Bigg[ \sum_{t=k}^{\widetilde{\ell}} \omega_{d-1-2k}^{i(t-k)} |t\rangle_A |00\rangle_{ab} + \omega_{d-1-2k}^{i(\widetilde{\ell}+1-k )} |\widetilde{\ell}+1 \rangle_A |11\rangle_{ab} + \cdots \\
	&\quad\quad\quad\quad\quad \quad +\omega_{d-1-2k}^{i(d-2-2k)}|d-2-k\rangle_A \left| (\ell -2-k)(\ell -2-k) \right\rangle_{ab}\Bigg] \left| \xi_j^{(d-2k)} \,\circ\, k \right\rangle_{\widetilde{B}}\Bigg\}.\\
\end{aligned}
\]
If $d$ is even, then additionally
\[
\mathcal{A}^{(d,0)} \to \left\{|\phi_r\rangle_A \left| \phi_s \circ \phi_t \right\rangle_{\widetilde{B}} |00\rangle_{ab}\right\}.
\]

\paragraph*{Step~2 (Bob).}
Bob performs the POVM
\[
\mathcal{M}_{2} \equiv \Big\{  M^{0}_{21}, \ldots, M^{\ell -2}_{21},\;
M^{0}_{22}, \ldots, M^{\ell-2}_{22},\;
M^{0}_{23}, \ldots, M^{\ell -2}_{23},\;
M_{24} \Big\}.
\]
Since the supports of these projectors are mutually orthogonal, the operator
\[
M_{24} := I_{\widetilde{B}b} - \sum_{m = 1}^3 \sum_{k = 0}^{\ell  - 2} {M_{2m}^k},
\]
is positive semidefinite. And
\[
\begin{aligned}
	M^{k}_{21} & = P\left[ \left| (|k+1\rangle,\dots, |d-1-k\rangle) \circ (|k\rangle,\dots, |d-2-k\rangle) \right\rangle_{\widetilde{B}} ; \left| \ell-1-k \right\rangle_{b} \right],\\
	M^{k}_{22} & = P\left[ \left|(d-1-k) \circ (|k+1\rangle,\dots, |d-1-k\rangle) \right\rangle_{\widetilde{B}} ; \left (|0\rangle, |1\rangle,\ldots,\left |\ell -2-k\right \rangle\right)_{b}  \right],\\
	M^{k}_{23} & = P\left[ \left|(|k+1\rangle,\dots, |d-1-k\rangle)  \circ k \right\rangle_{\widetilde{B}} ;\left (|0\rangle, |1\rangle,\ldots,\left |\ell -2-k\right \rangle\right )_{b} \right],
\end{aligned}
\]
for $k = 0,1,\ldots,\ell-2$.
The outcomes corresponding to $M^{k}_{21}$, $M^{k}_{22}$, and $M^{k}_{23}$ identify the subsets
$\mathcal{A}_3^{(d,d-2k)}$, $\mathcal{B}_1^{(d,d-2k)}$, and $\mathcal{B}_2^{(d,d-2k)}$, respectively.
If $M_{24}$ occurs, the remaining possibilities are
$\mathcal{A}_1^{(d,d-2k)}$, $\mathcal{A}_2^{(d,d-2k)}$, and $\mathcal{B}_3^{(d,d-2k)}$ (and, when $d$ is even, also $\mathcal{A}^{(d,0)}$).

\paragraph*{Step~3 (Alice).}
Alice performs
\[
\mathcal{M}_{3} \equiv \Bigg\{  M^{0}_{31}, \ldots, M^{\ell-2}_{31},\;
M_{32} := I_{Aa} - \sum_{k = 0}^{\ell   - 2} {M_{31}^k} \Bigg\},
\]
where $M^{k}_{31}=P[|k\rangle_{A};|0\rangle_{a}]$.
If $M^{k}_{31}$ clicks, the subset is $\mathcal{B}_3^{(d,d-2k)}$.
Otherwise $M_{32}$ clicks, leaving $\mathcal{A}_1^{(d,d-2k)}$ and $\mathcal{A}_2^{(d,d-2k)}$ (and, when $d$ is even, possibly $\mathcal{A}^{(d,0)}$).

\paragraph*{Step~4 (Bob).}
Bob performs
\[
\mathcal{M}_4 \equiv \Big\{ M^0_{41}, \ldots, M^{\ell -2}_{41},\;
M^{0}_{42}, \ldots, M^{\ell -2}_{42},\;
M_{43} \Big\},
\]
with
\[
M_{43} := I_{\widetilde{B}b}-\sum_{m=1}^{2}\sum_{k = 0}^{\ell  - 2} {M_{4m}^k},
\]
and
\[
\begin{aligned}
M^{k}_{41} &= P\left[ \left| k \circ \left( |k\rangle,|k+1\rangle,\dots,|d-k-2\rangle \right) \right\rangle_{\widetilde{B}};\;
\left( |0\rangle,\dots,\left|\ell  -1-k \right\rangle \right)_{b} \right],\\
M^{k}_{42} &= P\left[ \left| \left( |k\rangle,|k+1\rangle,\dots,|d-k-2\rangle \right) \circ (d-1-k) \right\rangle_{\widetilde{B}};\;
\left( |0\rangle,\dots,\left|\ell  -1-k \right\rangle \right)_{b} \right].
\end{aligned}
\]
If $M^{k}_{41}$ (resp.\ $M^{k}_{42}$) clicks, the subset is $\mathcal{A}_{1}^{(d,d-2k)}$ (resp.\ $\mathcal{A}_{2}^{(d,d-2k)}$).
The remaining outcome $M_{43}$ depends on the parity of $d$: if $d$ is odd, this branch is empty; if $d$ is even, it corresponds precisely to $\mathcal{A}^{(d,0)}$.

\medskip
\noindent
This completes the proof for the branch $M_{11}$. Since the other outcomes $M_{1m}$ lead to the same separation pattern by symmetry, the protocol perfectly identifies all subsets and hence yields perfect LOCC discrimination of $\mathcal{U}_d$ with the stated resources. \hfill $\square$

%================================================
\section{Proof of Theorem~5}\label{B}
%================================================
\paragraph*{Overview and resources.}
We follow the resource configuration and operational idea used in Theorem~1 (teleportation together with EPR pairs that are consumed sequentially according to the identified shell). Charlie and Bob share $|\phi^+(d)\rangle_{bc}$ and Charlie teleports $C$ to Bob, producing the bipartite state $|\psi\rangle_{A\widetilde{B}}$.

\subsection*{Round $m=0$ (outermost layer)}
\paragraph*{Step~$1^{(0)}$.}
Alice and Bob share one EPR pair $|\phi^+(2)\rangle_{a_0b_0}$.

\paragraph*{Step~$2^{(0)}$ (Alice).}
Alice performs
\[
\mathcal{M}^{0}_{1}\equiv \bigl\{M^{0}_{11},M^{0}_{12}\bigr\},
\]
where
\[\begin{aligned}
&M^{0}_{11}:= P\left[(|0\rangle,|1\rangle,\dots,|d-2\rangle)_A;|0\rangle_{a_0}\right] + P\left[|d-1\rangle_A;|1\rangle_{a_0}\right],\\
&M^{0}_{12}:= I - M^{0}_{11}.
\end{aligned}\]
Conditioned on $M^{0}_{11}$, the post-measurement states are exactly
\[
\begin{aligned}
	&\mathcal{A}_1^{(d,d)} \to \left\{ \left[ \left( \sum_{t=0}^{d-3} \omega_{d-1}^{jt} |t+1\rangle \right)_A |00\rangle_{a_0b_0} + \omega_{d-1}^{j(d-2)} |d-1\rangle_A |11\rangle_{a_0b_0} \right] |0 \circ \eta_i^{(d)} \rangle_{\widetilde{B}} \right\}, \\
	&\mathcal{A}_2^{(d,d)} \to \left\{ \left[ \left( \sum_{t=0}^{d-3} \omega_{d-1}^{jt} |t+1\rangle \right)_A |00\rangle_{a_0b_0} + \omega_{d-1}^{j(d-2)} |d-1\rangle_A |11\rangle_{a_0b_0} \right] |\eta_i^{(d)} \circ (d-1) \rangle_{\widetilde{B}} \right\}, \\
	&\mathcal{A}_3^{(d,d)} \to \left\{ |d-1\rangle_A |11\rangle_{a_0b_0} |\xi_j^{(d)} \circ \eta_i^{(d)} \rangle_{\widetilde{B}} \right\}, \\
	&\mathcal{B}_1^{(d,d)} \to \left\{ |\eta_i^{(d)} \rangle_A |00\rangle_{a_0b_0} |(d-1) \circ \xi_j^{(d)} \rangle_{\widetilde{B}} \right\}, \\
	&\mathcal{B}_2^{(d,d)} \to \left\{ |\eta_i^{(d)} \rangle_A |00\rangle_{a_0b_0} |\xi_j^{(d)} \circ 0 \rangle_{\widetilde{B}} \right\}, \\
	&\mathcal{B}_3^{(d,d)} \to \left\{ |0\rangle_A |00\rangle_{a_0b_0} |\eta_i^{(d)} \circ \xi_j^{(d)} \rangle_{\widetilde{B}} \right\}.\\
\end{aligned}
\]
Moreover, for $k=1,2,\ldots,\ell-2$,
\[
\begin{aligned}
	&\mathcal{A}_1^{(d,d-2k)} \to \left\{ |\xi_j^{(d-2k)} \rangle_A |00\rangle_{a_0b_0}|k \circ \eta_i^{(d-2k)} \rangle_{\widetilde{B}} \right\}, \\
	&\mathcal{A}_2^{(d,d-2k)} \to \left\{ |\xi_j^{(d-2k)} \rangle_A |00\rangle_{a_0b_0}|\eta_i^{(d-2k)} \circ (d-1-k) \rangle_{\widetilde{B}} \right\},\\
	&\mathcal{A}_3^{(d,d-2k)} \to \left\{ |d-1-k \rangle_A |00\rangle_{a_0b_0} |\xi_j^{(d-2k)} \circ \eta_i^{(d-2k)} \rangle_{\widetilde{B}} \right\}, \\
	&\mathcal{B}_1^{(d,d-2k)} \to \left\{ |\eta_i^{(d-2k)} \rangle_A |00\rangle_{a_0b_0} |(d-1-k) \circ \xi_j^{(d-2k)} \rangle_{\widetilde{B}} \right\}, \\
	&\mathcal{B}_2^{(d,d-2k)} \to \left\{ |\eta_i^{(d-2k)} \rangle_A |00\rangle_{a_0b_0} |\xi_j^{(d-2k)} \circ k \rangle_{\widetilde{B}} \right\}, \\
	&\mathcal{B}_3^{(d,d-2k)} \to \left\{ |k \rangle_A |00\rangle_{a_0b_0} |\eta_i^{(d-2k)} \circ \xi_j^{(d-2k)} \rangle_{\widetilde{B}} \right\}.
\end{aligned}
\]
If $d$ is even, then additionally
\[
\mathcal{A}^{(d,0)} \to \left\{|\phi_r\rangle_A \left| \phi_s \circ \phi_t \right\rangle_{\widetilde{B}} |00\rangle_{a_0b_0}\right\}.
\]

\paragraph*{Step~$3^{(0)}$ (Bob).}
Bob performs
\[
\begin{aligned}
	\mathcal{M}^{0}_{2} \equiv \bigl\{&
M^{0}_{21} := P\left[ |(1,\dots,d-1) \circ (0,\dots,d-2)\rangle_{\widetilde{B}} ; |1\rangle_{b_0} \right],\;
M^{0}_{22} := P\left[ |(d-1) \circ (1,\dots,d-1) \rangle_{\widetilde{B}} ; |0\rangle_{b_0} \right], \\
& M^{0}_{23} := P\left[ |(1,\dots,d-1) \circ 0 \rangle_{\widetilde{B}} ; |0\rangle_{b_0} \right],\;
M^{0}_{24} := I - M^{0}_{21} - M^{0}_{22} - M^{0}_{23} \bigr\}.
\end{aligned}
\]
Outcomes of $M^{0}_{21}$, $M^{0}_{22}$, $M^{0}_{23}$ identify $\mathcal{A}^{(d,d)}_{3}$, $\mathcal{B}^{(d,d)}_{1}$, and $\mathcal{B}^{(d,d)}_{2}$, respectively.
If $M^{0}_{24}$ clicks, the remaining possibilities are $\mathcal{A}^{(d,d)}_{1}$, $\mathcal{A}^{(d,d)}_{2}$, $\mathcal{B}^{(d,d)}_{3}$, and all deeper-layer subsets $\mathcal{P}^{(d,d-2k)}_{l}$ ($\mathcal{P}=\mathcal{A},\mathcal{B}$; $l=1,2,3$; $k\ge 1$), and (when $d$ is even) also $\mathcal{A}^{(d,0)}$.

\paragraph*{Step~$4^{(0)}$ (Alice).}
Alice performs
\[
\mathcal{M}^{0}_{3} \equiv \left\{
M^{0}_{31} := P\left[|0\rangle_A;|0\rangle_{a_0}\right],\;
M^{0}_{32}:= I  - M^{0}_{31}
\right\}.
\]
If $M^{0}_{31}$ clicks, the subset is $\mathcal{B}^{(d,d)}_{3}$.
If $M^{0}_{32}$ clicks, the remaining subsets are $\mathcal{A}^{(d,d)}_{1}$, $\mathcal{A}^{(d,d)}_{2}$, and deeper-layer subsets (and, when $d$ is even, possibly $\mathcal{A}^{(d,0)}$).

\paragraph*{Step~$5^{(0)}$ (Bob).}
Bob performs
\[
\begin{aligned}
	\mathcal{M}^{0}_{4} \equiv \bigl\{&
M^{0}_{41} := P\left[ |0 \circ (0,\dots,d-2)\rangle_{\widetilde{B}} ; I_{b_0} \right],\;
M^{0}_{42} := P\left[ |(0,\dots,d-2) \circ (d-1) \rangle_{\widetilde{B}} ; I_{b_0} \right],\\
& M^{0}_{43}:= I - M^{0}_{41} - M^{0}_{42} \bigr\}.
\end{aligned}
\]
Outcomes $M^{0}_{41}$ and $M^{0}_{42}$ identify $\mathcal{A}^{(d,d)}_{1}$ and $\mathcal{A}^{(d,d)}_{2}$, respectively (and these are LOCC-distinguishable).
If $M^{0}_{43}$ clicks, the remaining states lie entirely in deeper layers, and we proceed to the next round by consuming an additional EPR pair.

\subsection*{Iterating over inner layers}
The above five-step procedure isolates the outermost layer. We then add another EPR pair $|\phi^+(2)\rangle_{a_1b_1}$ and repeat the same logic on the remaining (inner-layer) subsets. The procedure is repeated for all shells $k=1,\ldots,\ell-2$, requiring a total of $\ell-1$ EPR pairs including the initial one.

\subsection*{Final round for odd $d$}
If $d$ is odd, after $\ell-2=\frac{d-3}{2}$ repetitions, the remaining subsets are exactly $\mathcal{P}^{(d,3)}_{l}$ ($\mathcal{P}=\mathcal{A},\mathcal{B}$; $l=1,2,3$).
Let $m=\frac{d-3}{2}$. We consume the $(m+1)$-th EPR pair $|\phi^+(2)\rangle_{a_{m}b_{m}}$ and perform the final discrimination procedure:

\paragraph*{Step~$1^{(m)}$.}
Alice and Bob share $|\phi^+(2)\rangle_{a_{m}b_{m}}$.

\paragraph*{Step~$2^{(m)}$ (Alice).}
Alice performs
\[
\mathcal{M}_{1}^{m} \equiv \Bigl\{
M_{11}^{m} := P\!\left[ \left( \left| \frac{d - 3}{2} \right\rangle , \left| \frac{d - 1}{2} \right\rangle \right)_A ; \left| 0 \right\rangle_{a_{m}} \right] + P\!\left[ \left| \frac{d + 1}{2} \right\rangle_A ; \left| 1 \right\rangle_{a_{m}} \right],\;
M_{12}^{m} := I - M_{11}^{m}
\Bigr\}.
\]
Conditioned on \(M_{11}^{m}\), the post-measurement states are
\[
\begin{aligned}
	&\mathcal{A}_1^{(d,3)} \to \Biggl\{
	\Biggl[ \left|\frac{d-1}{2}\right\rangle_A |00\rangle_{a_0b_0}\dots |00\rangle_{a_{m}b_{m}} + (-1)^{j} \left|\frac{d+1}{2}\right\rangle_A |00\rangle_{a_0b_0}\dots |11\rangle_{a_{m}b_{m}} \Biggr]
	\left|\frac{d-3}{2} \circ \eta_i^{(3)} \right\rangle_{\widetilde{B}} \Biggr\},\\
		&\mathcal{A}_2^{(d,3)} \to \Biggl\{
	\Biggl[ \left|\frac{d-1}{2}\right\rangle_A |00\rangle_{a_0b_0}\dots |00\rangle_{a_{m}b_{m}} +(-1)^{j} \left|\frac{d+1}{2}\right\rangle_A |00\rangle_{a_0b_0}\dots |11\rangle_{a_{m}b_{m}} \Biggr]
	\left|\eta_i^{(3)} \circ \frac{d+1}{2}\right\rangle_{\widetilde{B}} \Biggr\},\\
	&\mathcal{A}_3^{(d,3)} \to \left\{ \left |\frac{d+1}{2}\right \rangle_A |00\rangle_{a_0b_0}\dots |11\rangle_{a_{m}b_{m}} \left |\xi_j^{(3)} \circ \eta_i^{(3)} \right \rangle_{\widetilde{B}} \right\}, \\
	&\mathcal{B}_1^{(d,3)} \to \biggl\{
	\left|\eta_i^{(3)} \right\rangle_A |00\rangle_{a_0b_0}\dots |00\rangle_{a_{m}b_{m}}
	\left|\frac{d+1}{2}\circ \xi_j^{(3)} \right\rangle_{\widetilde{B}} \biggr\},\\
	&\mathcal{B}_2^{(d,3)} \to \biggl\{
	\left|\eta_i^{(3)} \right\rangle_A |00\rangle_{a_0b_0}\dots |00\rangle_{a_{m}b_{m}}
	\left|\xi_j^{(3)}\circ  \frac{d-3}{2}\right\rangle_{\widetilde{B}} \biggr\},\\
	&\mathcal{B}_3^{(d,3)} \to \biggl\{
	\left|\frac{d-3}{2} \right\rangle_A |00\rangle_{a_0b_0}\dots |00\rangle_{a_{m}b_{m}}
	\left|\eta_i^{(3)}\circ \xi_j^{(3)} \right\rangle_{\widetilde{B}} \biggr\}.
\end{aligned}
\]

\paragraph*{Step~$3^{(m)}$ (Bob).}
Bob performs
\[
\begin{aligned}
	\mathcal{M}^{m}_{2} \equiv \bigl\{ &
M^{m}_{21} := P\left[ \left |\left (\frac{d-1}{2},\frac{d+1}{2}\right) \circ \left (\frac{d-3}{2},\frac{d-1}{2}\right )\right \rangle_{\widetilde{B}} ; |1\rangle_{b_{m}} \right], \\
	& M^{m}_{22} := P\left[ \left|\frac{d+1}{2} \circ \left(\frac{d-1}{2},\frac{d+1}{2}\right) \right\rangle_{\widetilde{B}} ; |0\rangle_{b_{m}} \right],\\
	& M^{m}_{23} := P\left[\left  |\left (\frac{d-1}{2},\frac{d+1}{2}\right ) \circ \frac{d-3}{2} \right\rangle_{\widetilde{B}} ; |0\rangle_{b_{m}} \right], \\
	& M^{m}_{24} := I - M^{m}_{21} - M^{m}_{22} - M^{m}_{23} \bigr\}.
\end{aligned}
\]
Then $M^{m}_{21}$, $M^{m}_{22}$, and $M^{m}_{23}$ identify $\mathcal{A}_3^{(d,3)}$, $\mathcal{B}_1^{(d,3)}$, and $\mathcal{B}_2^{(d,3)}$, respectively; $M^{m}_{24}$ leaves $\mathcal{A}_1^{(d,3)}$, $\mathcal{A}_2^{(d,3)}$, and $\mathcal{B}_3^{(d,3)}$.

\paragraph*{Step~$4^{(m)}$ (Alice).}
Alice performs
\[
\mathcal{M}_3^{m} \equiv \left\{
M_{31}^{m} := P\left[\left| \frac{d-3}{2} \right\rangle_A ; |0\rangle_{a_{m}}\right],\;
M_{32}^{m} := I - M_{31}^{m}\right\}.
\]
Outcome $M^{m}_{31}$ identifies $\mathcal{B}^{(d,3)}_{3}$; otherwise $M^{m}_{32}$ leaves $\mathcal{A}^{(d,3)}_{1}$ and $\mathcal{A}^{(d,3)}_{2}$.

\paragraph*{Step~$5^{(m)}$ (Bob).}
Bob performs
\[
\mathcal{M}^{m}_{4} \equiv \left\{
M^{m}_{41} := P\left[\left|\frac{d-3}{2}\circ \left(\frac{d-3}{2},\frac{d-1}{2}\right)\right\rangle_{\widetilde{B}} ; I_{b_{m}} \right],\;
M^{m}_{42} := I - M^{m}_{41} \right\},
\]
so that $M^{m}_{41}$ and $M^{m}_{42}$ identify $\mathcal{A}^{(d,3)}_{1}$ and $\mathcal{A}^{(d,3)}_{2}$, respectively. Both are LOCC-distinguishable. This completes the proof for the odd-dimensional case.

\subsection*{Final discrimination for even $d$}
If $d$ is even, after $\ell-2=\frac{d-4}{2}$ repetitions, the remaining subsets are $\mathcal{A}^{(d,0)}$ and $\mathcal{P}^{(d,4)}_{l}$ ($\mathcal{P}=\mathcal{A},\mathcal{B}$; $l=1,2,3$).
Let $m=\frac{d-4}{2}$. The inner-layer subsets $\mathcal{A}^{(d,4)}_{3}$, $\mathcal{B}^{(d,4)}_{1}$, $\mathcal{B}^{(d,4)}_{2}$, and $\mathcal{B}^{(d,4)}_{3}$ are distinguished exactly as in the odd-dimensional case (with the corresponding parameters).
The final discrimination step separating $\mathcal{A}^{(d,4)}_{1}$, $\mathcal{A}^{(d,4)}_{2}$, and $\mathcal{A}^{(d,0)}$ is:

\paragraph*{Final Step (Bob).}
Bob performs
\[
\begin{aligned}
	\mathcal{M}^{m}_{\mathrm{fin}} \equiv \bigg\{
	& M^{m}_{\mathrm{fin},1} := P\left[\left|\frac{d-4}{2}\circ \left(\frac{d-4}{2}, \frac{d-2}{2}, \frac{d}{2}\right)\right\rangle_{\widetilde{B}} ; I_{b_{m}} \right], \\
	& M^{m}_{\mathrm{fin},2} := P\left[\left|\left(\frac{d-4}{2}, \frac{d-2}{2}, \frac{d}{2}\right)\circ \frac{d+2}{2}\right\rangle_{\widetilde{B}} ; I_{b_{m}} \right], \\
	& M^{m}_{\mathrm{fin},3} := I - M^{m}_{\mathrm{fin},1} - M^{m}_{\mathrm{fin},2}\bigg\},
\end{aligned}
\]
and the measurement outcomes satisfy
\[
M^{m}_{\mathrm{fin},1}\Rightarrow \mathcal{A}^{(d,4)}_{1},\qquad
M^{m}_{\mathrm{fin},2}\Rightarrow \mathcal{A}^{(d,4)}_{2},\qquad
M^{m}_{\mathrm{fin},3}\Rightarrow \mathcal{A}^{(d,0)}.
\]

\paragraph*{Entanglement consumption.}
The teleportation step consumes \(\log_2 d\) ebits. Assume that the
unknown state is chosen uniformly from \(\mathcal U_d\). The
\((m+1)\)-th EPR pair is consumed precisely when the protocol reaches
the \(m\)-th discrimination round.

Let \(N_m\) denote the number of candidate states remaining before the
\(m\)-th shell is processed. From the nested construction,
\[
N_m=(d-2m)^3-8(\ell-1-m),
\qquad
m=0,1,\ldots,\ell-2.
\]
The total number of states is
\[
N_{\mathrm{tot}}=d^3-8(\ell-1).
\]
Therefore, the probability of reaching the \(m\)-th round is
\[
p_m=\frac{N_m}{N_{\mathrm{tot}}}
=
\frac{(d-2m)^3-8(\ell-1-m)}
     {d^3-8(\ell-1)}.
\]
Hence the average number of consumed EPR pairs is
\[
e=\sum_{m=0}^{\ell-2}p_m
=
\sum_{m=0}^{\ell-2}
\frac{(d-2m)^3-8(\ell-1-m)}
     {d^3-8(\ell-1)}.
\]
Accordingly, the average bipartite entanglement consumption is
\[
\log_2 d+e
\]
ebits.\hfill $\square$

%================================================
\section{Proof of Theorem~6}\label{C}
%================================================
\paragraph*{Resources and initial state.}
We extend the teleportation-free idea of Theorem~2 to general $d$.
Alice shares a maximally entangled state of local dimension $\ell$ with Bob, and another with Charlie:
\[
|\phi^+(\ell)\rangle_{a_1b_1}
=\frac{1}{\sqrt{\ell}}\sum_{r=0}^{\ell-1}|rr\rangle_{a_1b_1},
\qquad
|\phi^+(\ell)\rangle_{a_2c_1}
=\frac{1}{\sqrt{\ell}}\sum_{r=0}^{\ell-1}|rr\rangle_{a_2c_1}.
\]
No teleportation is used.

\paragraph*{Step~1 (Bob and Charlie).}
Bob performs the POVM
\[
\mathcal{M}_{1}\equiv \bigl\{M_{11},M_{12},\ldots,M_{1\ell }\bigr\},
\]
where
\[
\begin{aligned}
\mathcal{M}_{1}\equiv \Big\{&M_{11}:=P[|0\rangle_B;|0\rangle_{b_1}]+P[|1\rangle_B;|1\rangle_{b_1}]+\cdots+P[\left(\left |\ell -1\right \rangle,\ldots,|d-1\rangle\right )_B;\left |\ell-1\right \rangle_{b_1}],\\
	&M_{12}:=P[|0\rangle_B;|1\rangle_{b_1}]+P[|1\rangle_B;|2\rangle_{b_1}]+\cdots+P [\left(\left|\ell -1\right\rangle,\ldots,|d-1\rangle\right)_B;|0\rangle_{b_1}],\\
	&\quad\quad\quad \vdots \\
	&M_{1\ell }:=P [|0\rangle_B;\left |\ell -1\right \rangle_{b_1}]+P[|1\rangle_B;|0\rangle_{b_1}]+\cdots+P [\left(\left|\ell -1\right \rangle,\ldots,|d-1\rangle\right)_B;\left|\ell-2\right\rangle_{b_1}]
	\Big \}.
\end{aligned}
\]
Charlie performs the projective measurement:
\[
\mathcal{M}_{2}\equiv \bigl\{M_{21},M_{22},\ldots,M_{2\ell}\bigr\},
\]
\[
\begin{aligned}
	\mathcal{M}_{2} \equiv \Bigl\{
	&	M_{21} := P \big[ (\vert0\rangle, \dots,  \vert  \widetilde{\ell}\rangle )_C; \vert0\rangle_{c_1}\big] + P \big[ \vert\widetilde{\ell}+ 1 \rangle_C; \vert1\rangle_{c_1}\big] + \cdots + P\big[\vert d-1\rangle_C; \left \vert \ell - 1\right \rangle_{c_1} \big], \\
	&	M_{22} := P\big [(\vert0\rangle, \dots,  \vert\widetilde{\ell} \rangle)_C; \vert1\rangle_{c_1}\big ] + P\big[ \vert\widetilde{\ell} + 1 \rangle_C; \vert2\rangle_{c_1}\big] + \cdots  + P\big [\vert d-1\rangle_C; \vert 0\rangle_{c_1}\big], \\
	&\quad\quad\quad \vdots \\
	&M_{2\ell } := P\big[(\vert0\rangle, \dots, \vert\widetilde{\ell}\rangle)_C;  \vert\ell  -1 \rangle_{c_1}\big] + P\big[  \vert\widetilde{\ell} + 1\rangle_C; \vert0\rangle_{c_1}\big] + \cdots \\
	&\quad \quad \quad\quad\quad + P\big[\vert d-1\rangle_C; \vert\ell -2 \rangle_{c_1}\big ]
	\Bigr\}.
\end{aligned}
\]

Conditioned on obtaining outcomes \(M_{11}\) and \(M_{21}\), the post-measurement states are as follows.
\[
\begin{aligned}
	&\mathcal{A}_1^{(d,d-2k)} \to \Bigg \{ \left|\xi_j^{(d-2k)}\right\rangle_A |k\rangle_B |kk\rangle_{a_1b_1} \Bigg [ \sum_{t=k}^{\widetilde{\ell}} \omega_{d-1-2k}^{i(t-k)} |t\rangle_C |00\rangle_{a_2c_1}  + \cdots \\
	&\quad\quad\quad\quad\quad \hspace{1em} +\omega_{d-1-2k}^{i(d-2-2k)} |d-2-k\rangle_C \left |\left(\ell -2-k\right) \left(\ell -2-k\right)\right \rangle_{a_2c_1} \Bigg]\Bigg  \},\\
	&\mathcal{A}_2^{(d,d-2k)} \to \Bigg \{ \left |\xi_j^{(d-2k)}\right\rangle_A\Bigg [ |k\rangle_B |kk\rangle_{a_1b_1} +\cdots+ \sum_{t=\ell -1}^{d-2-k} \omega_{d-1-2k}^{i(t-k)} |t\rangle_B\left |\left(\ell-1\right)\left(\ell -1\right)\right\rangle_{a_1b_1} \Bigg]\\
	&\quad \quad\quad\quad\quad\hspace{1.2em} |d-1-k\rangle_C \left|\left(\ell -1-k\right) \left(\ell -1-k\right)\right\rangle_{a_2c_1} \Bigg \},\\
	&\mathcal{A}_3^{(d,d-2k)} \to \Bigg\{ |d-1-k\rangle_A\Bigg [ |k+1\rangle_B |\left(k+1\right)\left(k+1\right)\rangle_{a_1b_1} + \cdots +\sum_{t=\ell -2}^{d-2-k} \omega_{d-1-2k}^{j(t-k)} |t+1\rangle_B\\
	&\quad \quad\quad\quad\quad \quad  \left|\left(\ell -1\right) \left(\ell  -1\right)\right\rangle_{a_1b_1} \Bigg]  \Bigg[ \sum_{t=k}^{\widetilde{\ell} } \omega_{d-1-2k}^{i(t-k)} |t\rangle_C |00\rangle_{a_2c_1} + \cdots \\
	&\quad \quad\quad\quad\quad \quad +\omega_{d-1-2k}^{i(d-2-2k)} |d-2-k\rangle_C \left |\left(\ell  -2-k\right) \left(\ell -2-k\right)\right \rangle_{a_2c_1} \Bigg ] \Bigg\}, \\
	&\mathcal{B}_1^{(d,d-2k)} \to \Bigg \{\left |\eta_i^{(d-2k)}\right\rangle_A |d-1-k\rangle_B \left|\left(\ell  -1\right) \left(\ell -1\right)\right\rangle_{a_1b_1}\Bigg [ \sum_{t=k}^{\widetilde{\ell} -1} \omega_{d-1-2k}^{j(t-k)} |t+1\rangle_C  \\
&  \quad \quad\quad\quad\quad \quad   \hspace{0.25em} |00\rangle_{a_2c_1}+ \cdots + \omega_{d-1-2k}^{j(d-2-2k)} |d-1-k\rangle_C \left|\left(\ell-1-k\right) \left(\ell-1-k\right)\right\rangle_{a_2c_1}\Bigg  ] \Bigg \}, \\
\end{aligned}
\]
	\[
	\begin{aligned}
&\mathcal{B}_2^{(d,d-2k)} \to\Bigg  \{\left  |\eta_i^{(d-2k)}\right\rangle_A \Bigg[ |k+1\rangle_B |\left(k+1\right) \left(k+1\right)\rangle_{a_1b_1} + \cdots  + \sum_{t=\ell-2}^{d-2-k} \omega_{d-1-2k}^{j(t-k)} |t+1\rangle_B\\
&\quad \quad\quad\quad\quad \quad   \left|\left(\ell-1\right) \left(\ell-1\right)\right\rangle_{a_1b_1} \Bigg] |k\rangle_C |00\rangle_{a_2c_1} \Bigg\},\\
	&\mathcal{B}_3^{(d,d-2k)} \to \Bigg\{ |k\rangle_A\Bigg  [ |k\rangle_B |kk\rangle_{a_1b_1} +\cdots  + \sum_{t=\ell-1}^{d-2-k} \omega_{d-1-2k}^{i(t-k)} |t\rangle_B \left|\left(\ell-1\right) \left(\ell-1\right)\right\rangle_{a_1b_1} \Bigg] \\
		&\quad \quad\quad\quad\quad \quad   \Bigg[ \sum_{t=k}^{\widetilde{\ell} -1} \omega_{d-1-2k}^{j(t-k)} |t+ 1\rangle_C |00\rangle_{a_2c_1} + \cdots + \omega_{d-1-2k}^{j(d-2-2k)} |d-1-k\rangle_C \\
			&\quad \quad\quad\quad\quad \quad  \left|\left(\ell-1-k\right) \left(\ell-1-k\right)\right\rangle_{a_2c_1} \Bigg] \Bigg\}.
\end{aligned}
\]

If $d$ is even, then additionally
\[
\mathcal{A}^{(d,0)} \to\left\{\left|\phi_r \right\rangle_A \left|\phi_s\right \rangle_B \left|\phi_t \right\rangle_C\left| \left(\ell-1\right) \left(\ell-1\right)\right\rangle_{a_1b_1}|00\rangle_{a_2c_1}\right\}.
\]

\paragraph*{Step~2 (Alice).}
Alice performs the POVM $\mathcal{M}_3$ defined by
\[
\mathcal{M}_{3} \equiv \Big\{
M^{0}_{31}, \ldots, M^{\ell -2}_{31},\;
M^{0}_{32}, \ldots, M^{\ell -2}_{32},\;
M_{33} \Big\},
\]
where
\[M_{33}
=
I-
\sum_{k=0}^{\ell-2}
\left(M_{31}^{k}+M_{32}^{k}\right),\]
and
\[
\begin{aligned}
	M^{k}_{31} &:= P\big[\left(|k+1\rangle,\dots,|d-1-k\rangle\right)_{A};|k\rangle_{a_1};\left(|0\rangle,\dots,\left|\ell-2-k\right\rangle\right)_{a_2}\big],\\
	M^{k}_{32} &:= P\big[|d-1-k\rangle_{A};\left(|k+1\rangle,\dots,\left|\ell-1\right\rangle\right)_{a_1};\left(|0\rangle,\dots,\left|\ell-2-k\right\rangle\right)_{a_2}\big].
\end{aligned}
\]
These outcomes identify
\[
M^{k}_{31} \Rightarrow \mathcal{A}_1^{(d,d-2k)},\qquad
M^{k}_{32} \Rightarrow \mathcal{A}_3^{(d,d-2k)},
\]
while $M_{33}$ leaves $\mathcal{A}_2^{(d,d-2k)}$ or $\mathcal{B}_{l}^{(d,d-2k)}$ ($l=1,2,3$) (and, if $d$ is even, possibly also $\mathcal{A}^{(d,0)}$). In the following we consider the branch corresponding to $M_{33}$.

\paragraph*{Step~3 (Charlie).}
Charlie performs
\[
\mathcal{M}_{4}\equiv \Big\{
M^{0}_{41},\ldots,M^{\ell -2}_{41},\;
M_{42}\Big\},
\qquad
M_{42}=I-\sum_{k=0}^{\ell-2}M^{k}_{41},
\]
where $M^{k}_{41}:=P[|k\rangle_{C};|0\rangle_{c_1}]$.
If $M^{k}_{41}$ clicks, the subset is $\mathcal{B}_{2}^{(d,d-2k)}$.
Otherwise ($M_{42}$), the remaining subsets are $\mathcal{A}_2^{(d,d-2k)}$, $\mathcal{B}_{1}^{(d,d-2k)}$, and $\mathcal{B}_{3}^{(d,d-2k)}$ (and, when $d$ is even, possibly $\mathcal{A}^{(d,0)}$).

\paragraph*{Step~4 (Bob).}
Bob performs
\[
\mathcal{M}_{5}\equiv \Big\{
M^{0}_{51},\ldots,M^{\ell-2}_{51},\;
M_{52}\Big\},
\qquad
M_{52}=I-\sum_{k=0}^{\ell-2}M^{k}_{51},
\]
where
\[
M^{k}_{51}:=P\left[|d-1-k\rangle_{B};\left|\ell-1\right\rangle_{b_1}\right].
\]
If $M^{k}_{51}$ clicks, the subset is $\mathcal{B}_{1}^{(d,d-2k)}$; otherwise, the remaining subsets are $\mathcal{A}_2^{(d,d-2k)}$ and $\mathcal{B}_{3}^{(d,d-2k)}$ (and, when $d$ is even, possibly $\mathcal{A}^{(d,0)}$).

\paragraph*{Step~5 (Alice).}
Finally, Alice performs
\[
\mathcal{M}_{6} \equiv \Big\{
M^{0}_{61}, \ldots, M^{\ell -2}_{61},\;
M^{0}_{62}, \ldots, M^{\ell -2}_{62},\;
M_{63}\Big\},
\]
where
\[M_{63}
=
I-
\sum_{k=0}^{\ell-2}
\left(M_{61}^{k}+M_{62}^{k}\right),\]
and
\[
\begin{aligned}
M^{k}_{61} &:= P\left [(|k+1\rangle,\dots,|d-1-k\rangle)_{A};\left(|k\rangle,\dots,\left |\ell-1\right\rangle \right)_{a_1};\left |\ell-1-k\right\rangle_{a_2}\right],\\
M^{k}_{62} &:= P\left [|k\rangle_{A};\left(|k\rangle,\dots,\left|\ell -1\right\rangle\right )_{a_1};\left(|0\rangle,\dots,\left|\ell-1-k\right\rangle\right)_{a_2}\right ].
\end{aligned}
\]
Outcomes $M^{k}_{61}$ and $M^{k}_{62}$ identify $\mathcal{A}_{2}^{(d,d-2k)}$ and $\mathcal{B}_{3}^{(d,d-2k)}$, respectively.
The remaining outcome $M_{63}$ depends on the parity of $d$: if $d$ is odd, this branch is empty; if $d$ is even, it corresponds exactly to $\mathcal{A}^{(d,0)}$.

\paragraph*{Conclusion and resource cost.}
All subsets are therefore deterministically identified by LOCC. The protocol requires two Schmidt-rank-$\ell$ maximally entangled states throughout the discrimination process. Under the resource-accounting convention adopted in this paper, the total supplied bipartite entanglement is therefore \( 2\log_2\ell \) ebits, as stated in Theorem~6. \hfill $\square$
\end{appendix}

\end{document}